\documentclass[12pt,a4paper]{article}
\usepackage{a4wide}
\usepackage{amsmath}
\usepackage{amssymb}
\usepackage{epsfig}
\usepackage{subfigure}
\usepackage{exscale}
\usepackage{float}
\usepackage{bbm}
\usepackage[numbers,sort&compress]{natbib}
\newcommand{\N}{{\mathbb{N}}}
\newcommand{\Z}{{\mathbb{Z}}}

\newcommand{\p}{\partial}

\newcommand{\ontopof}[2]{\genfrac{}{}{0pt}{}{#1}{#2}}

\makeatletter
\@addtoreset{equation}{section}
\makeatother

\title{Holes Localized on a Skyrmion in a Doped Antiferromagnet on the
Honeycomb Lattice: Symmetry Analysis}

\author{N.\ D.\ Vlasii$^{a,d}$, C.\ P.\ Hofmann$^b$, F.-J.\ Jiang$^c$, and
U.-J.\ Wiese$^d$
\\ \\
$^a$ Bogolyubov Institute for Theoretical Physics,\\
National Academy of Sciences of Ukraine,\\
14-b Metrologichna Str., Kyiv, 03680, Ukraine \\ \\
$^b$ Facultad de Ciencias, Universidad de Colima \\
Colima C.P.\ 28045, Mexico \\ \\
$^c$ Department of Physics, National Taiwan Normal University \\
88, Sec.\ 4, Ting-Chou Rd., Taipei 116, Taiwan \\ \\
$^d$ Albert Einstein Center for Fundamental Physics \\
Institute for Theoretical Physics, Bern University \\
Sidlerstrasse 5, CH-3012 Bern, Switzerland}

\begin{document}

\maketitle

\begin{abstract} \normalsize

Using the low-energy effective field theory for hole-doped antiferromagnets on
the honeycomb lattice, we study the localization of holes on Skyrmions, as a 
potential mechanism for the preformation of Cooper pairs. In contrast to the
square lattice case, for the standard radial profile of the Skyrmion on the
honeycomb lattice, only holes residing in one of the two hole pockets can get 
localized. This differs qualitatively from hole pairs bound by magnon exchange,
which is most attractive between holes residing in different momentum space
pockets. On the honeycomb lattice, magnon exchange unambiguously leads to 
$f$-wave pairing, which is also observed experimentally. Using the 
collective-mode quantization of the Skyrmion, we determine the quantum numbers 
of the localized hole pairs. Again, $f$-wave symmetry is possible, but other 
competing pairing symmetries cannot be ruled out.

\end{abstract}

\maketitle

\section{Introduction}

Understanding the mechanism underlying high-temperature superconductivity has
remained a major challenge in condensed matter physics. Since high-temperature
cuprate superconductors are insulating antiferromagnets before doping, it is
natural to also investigate their antiferromagnetic precursors. In particular,
one may hope to identify potential mechanisms for Cooper pair preformation in
the antiferromagnetic phase. While we do not necessarily expect to unravel the
relevant mechanism in this way, it motivates a careful systematic study. In 
previous work we have investigated the interactions between holes in lightly 
doped antiferromagnets, using a systematic low-energy effective field theory 
approach, both on the square \citep{Bru06} and on the honeycomb lattice
\cite{Kae11}.

The effective theory is formulated in terms of the staggered magnetization
order parameter field, whose  fluctuations correspond to spinwaves (magnons),
and in terms of fermionic hole fields. This is in complete analogy to baryon
chiral perturbation theory in particle physics, where the fluctuations in the
chiral order parameter manifest themselves as pions, while baryons (protons and
neutrons) are analogous to the doped holes \citep{Gas88,Jen91,Ber92,Bec99}.
Based on microscopic Hubbard and $t$-$J$ models, and using a systematic
low-energy expansion, we have constructed effective field theories for magnons
and holes, both on the square and on the honeycomb lattice
\citep{Kae05,Bru05,Bru06,Bru07,Bru07a,Jia09,Kae11}.
In both cases, one-magnon exchange mediates weak attractive forces between
doped holes. As doping is increased, antiferromagnetism is weakened, and
ultimately disappears as a long-range order phenomenon, when one enters the
superconducting phase. Still, intermediate-range antiferromagnetic
correlations persist even in the superconductor, and it is interesting to ask
which objects form when one is about to leave the antiferromagnetic phase. At
the edge of this phase, the spin stiffness $\rho_s$ decreases and the energy
$4 \pi \rho_s$ of Skyrmion excitations in the staggered magnetization order
parameter is thus reduced. In addition, doped holes can gain energy when they
localize on a topological Skyrmion defect. We have systematically investigated
this phenomenon as a potential mechanism for Cooper pair preformation for
antiferromagnets on the square lattice \citep{VHJW12}. Interestingly, in this
case, both one-magnon exchange and Skyrmion localization lead to bound states 
in the same symmetry channel. The role of Skyrmions in quantum antiferromagnets 
has been investigated in \citep{Hal88,Rea89,Shr90,Goo91,Goo93,Haa96,Mar01,Mot03,Sen04,Bae04,Wie05,Mor05,Fu10,Rai11,Bas11,Ver91,Sei98,Ber99,Tim00}.

The main purpose of the present paper is to extend the study of hole
localization on a Skyrmion to antiferromagnets on the honeycomb lattice, which
underlies certain high-temperature superconductors, including the
dehydrated version of Na$_2$CoO$_2 \times y$H$_2$O. In this case, experiments
suggest that the pairing symmetry is $f$-wave \citep{MJ05}. $F$-wave pairing 
has also been found for other strongly correlated systems on the honeycomb
lattice \citep{Hon08,Lee10,Gan13}. As we studied earlier \citep{Kae11}, in 
contrast to the square lattice case, on the honeycomb lattice, leading order 
one-magnon exchange gives rise to long-range attraction only between holes 
residing in different hole pockets (characterized by the ``flavors'' $\alpha$ 
and $\beta$). As an unambiguous prediction of the effective theory, the binding 
occurs in the $f$-wave channel, and is thus indeed consistent with experiment
\cite{MJ05}. As we will show here, unlike in the square lattice case, on the 
honeycomb lattice the symmetry channels favored by hole localization on 
Skyrmions are not in one-to-one correspondence with the symmetry channels 
resulting from one-magnon exchange. In particular, a Skyrmion with the standard 
radial profile can only localize holes residing in the $\alpha$-pocket, while 
an anti-Skyrmion can only localize $\beta$-holes. While $f$-wave symmetry
still arises, other competing pairing symmetries are possible as well. Only
detailed energetic considerations, which we leave for future work, can 
unambiguously decide which pairing mechanism is favored by Skyrmion 
localization. In this paper, we concentrate entirely on the systematic symmetry 
analysis of the various hole states localized on a Skyrmion.

The rest of the paper is organized as follows. In Section 2 we review the
effective field theory formulation of antiferromagnetic magnons on the
honeycomb lattice and discuss Skyrmions as classical solutions. We also comment
on the Hopf term and on the collective modes of a rotating Skyrmion which we
then quantize. In Section 3, the effective field theory description is extended
by introducing holes doped into the system. Section 4 is devoted to the
symmetry analysis of states of single holes as well as hole pairs of the same
flavor, localized on a static or a rotating Skyrmion. While our main focus is
the symmetry analysis, some simple energetic considerations are also discussed.
While Section 6 contains our conclusions, the localization of two holes with
different flavor on a Skyrmion is investigated in the Appendix.

\section{Skyrmions in Magnon Effective Field Theory}

In this section, we consider the effective description of magnons and
Skyrmions, i.e., we restrict ourselves to the undoped honeycomb lattice
antiferromagnet.

\subsection{Symmetries of the Effective Action}

Magnons are the Goldstone bosons which result from the spontaneously broken
spin symmetry $SU(2)_s \to U(1)_s$. They are described by a 3-component
unit-vector field ${\vec e}(x) \in S^2$, living in the coset space
$S^2 = SU(2)_s/U(1)_s$. The coordinate $x = (x_1,x_2,t)$ represents a point in
$(2+1)$-dimensional Euclidean space-time, while the vector $\vec e(x)$ points
into the direction of the local staggered magnetization, i.e., into the
direction of the order parameter of the spontaneously broken spin rotation
symmetry. The Euclidean effective action for the magnons, at leading order in
the systematic derivative expansion, takes the form
\begin{equation}
\label{effact}
S[\vec e] = \int d^2x \ dt \ \frac{\rho_s}{2}
\left( \p_i \vec e \cdot \p_i \vec e +
\frac{1}{c^2} \p_t \vec e \cdot \p_t \vec e \right) .
\end{equation}
The low-energy couplings $\rho_s$ and $c$ are the spin stiffness and the
spinwave velocity, respectively. The ground state of the system is described
by a constant staggered magnetization vector which we choose to point in the
$3$-direction: $\vec e(x) = (0,0,1)$. The spin waves or magnons then correspond
to small fluctuations around the vector $\vec e(x) = (0,0,1)$. Unlike for
ferromagnetic magnons which follow a quadratic dispersion law,
antiferromagnetic magnons obey a linear, i.e., ``relativistic'' dispersion
relation. Note that the leading-order effective action for the magnons on the
honeycomb lattice, Eq.~(\ref{effact}), is identical with the one on the square
lattice. In general, lattice anisotropies only start manifesting themselves at
subleading orders.

Note that the $SU(2)_s$ spin symmetry corresponds to an internal symmetry, much
like chiral symmetry in particle physics. Hence its unbroken $U(1)_s$ subgroup
(which in particle physics corresponds to isospin) also represents an internal
symmetry. In view of this analogy, transformations in the unbroken subgroup
$U(1)_s$ are denoted by $I(\gamma)$. In the construction of the effective field
theory, the spontaneously broken spin symmetry $SU(2)_s \to U(1)_s$ plays an
essential role. In particular, global transformations in the unbroken subgroup
$U(1)_s$ will be important. We parametrize the order parameter vector as
\begin{equation}
\vec e(x) =
(\sin \theta(x) \cos \varphi(x),\sin \theta(x) \sin \varphi(x), \cos\theta(x)) .
\end{equation}
Under the global transformations, this vector transforms as
\begin{equation}
\label{subgroup}
^{I(\gamma)}\vec e(x) = (\sin\theta(x) \cos(\varphi(x) + \gamma),
\sin\theta(x) \sin(\varphi(x) + \gamma),\cos\theta(x)) .
\end{equation}

\begin{figure}
\begin{center}
\includegraphics[scale=0.7]{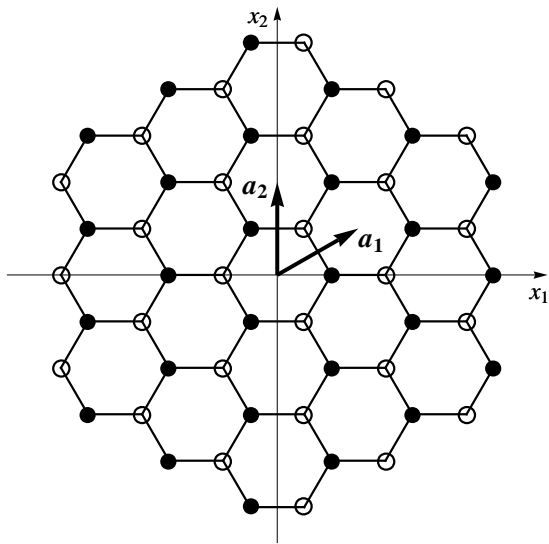} \vspace{-1.0cm}
\end{center}
\caption{\it The bipartite non-Bravais honeycomb lattice consists of two
triangular Bravais sublattices. The quantities $a_1$ and $a_2$ are the two
independent primitive lattice vectors.}
\label{honeycombLattice}
\end{figure}

Apart from the spontaneously broken spin symmetry $SU(2)_s$, the effective
action exhibits further symmetries, both continuous and discrete. First of all,
the leading-order expression of the effective action, Eq.~(\ref{effact}), is
Poin\-ca\-r\'e invariant. Note that we are dealing with an accidental symmetry
on the effective level, which is not shared by the underlying Hubbard or
$t$-$J$ models. It only emerges at leading order of the effective action. The
discrete symmetries are translations, rotations, and reflections of the
underlying honeycomb lattice, which we have depicted in
Fig.~\ref{honeycombLattice} with its two translation vectors $a_1$ and $a_2$.
The displacements $D_i$ along these primitive lattice vectors leave the
staggered magnetization invariant, such that the field $\vec e(x)$ transforms
trivially,
\begin{equation}
^{D_i}\vec e(x) = \vec e(x).
\end{equation}
Rotations by 60 degrees around an axis located at the center of a hexagon act
on the staggered magnetization vector as
\begin{equation}
^{O}\vec e(x) = - \vec e(O x).
\end{equation}
Note that, on the honeycomb lattice the discrete rotation $O$ by 60 degrees is 
spontaneously broken, while on the square lattice rotations by 90 degrees are 
not. It is convenient to also introduce the modified rotation symmetry, $O'$, 
which is a combination of the simple rotation $O$ and the $SU(2)_s$ spin 
rotation $g = i \sigma_2$. In contrast to $O$, the combined symmetry $O'$ is not
spontaneously broken. Under $O'$ the staggered magnetization vector transforms 
as
\begin{equation}
^{O'} \vec e(x) = (e_1(Ox),- e_2(Ox),e_3(Ox)) .
\end{equation}
Finally, we have to consider spatial reflections at the $x_1$-axis
(see Fig.~\ref{honeycombLattice}) which act as
\begin{equation}
^R\vec e(x) = \vec e(Rx), \qquad
Rx = (x_1,-x_2,t) = (r \cos\chi,- r \sin\chi,t) ,
\end{equation}
as well as time reversal, which changes the direction of a spin, and is
represented by
\begin{equation}
^T\vec e(x) = - \vec e(Tx), \qquad
Tx = (x_1,x_2,-t) = (r \cos\chi,r \sin\chi,-t) .
\end{equation}
The effective action, Eq.~(\ref{effact}), respects all these symmetries which
(except for the accidental Poin\-ca\-r\'e invariance) are inherited from the
underlying Hubbard or $t$-$J$ models.

\subsection{Skyrmion Solutions}

Skyrmions are topologically non-trivial classical solutions of the magnon
effective field theory. Their integer winding number
\begin{equation}
n[\vec e] = \frac{1}{8 \pi} \int d^2x \ \varepsilon_{ij} \vec e \cdot
\left[\p_i \vec e \times \p_j \vec e\right] \in \Pi_2[S^2] = \Z,
\end{equation}
takes values in the second homotopy group of the order parameter sphere $S^2$.
The topological current
\begin{equation}
j_\mu(x) = \frac{1}{8 \pi} \varepsilon_{\mu\nu\rho} \vec e(x) \cdot
\left[\p_\nu \vec e(x) \times \p_\rho \vec e(x)\right],
\end{equation}
represents a conserved quantity, i.e.\ $\p_\mu j_\mu(x) = 0$. The time component
of the topological current is related to the winding number by
$n[\vec e] = \int d^2x \ j_t(x)$ and thus represents the integrated topological
charge density. The transformation properties of the topological charge density
with respect to the relevant symmetries are
\begin{eqnarray}
U(1)_s:&&^{I(\gamma)}j_t(x) = j_t(x), \nonumber \\
D_i:&&^{D_i}j_t(x) = j_t(x), \nonumber \\
O:&&^{O}j_t(x) = - j_t(O x), \nonumber \\
O':&&^{O'}j_t(x) = - j_t(O x), \nonumber \\
R:&&^Rj_t(x) = - j_t(Rx), \nonumber \\
T:&&^Tj_t(x) = - j_t(Tx).
\end{eqnarray}
The winding number thus picks up a sign under the rotations $O$ and $O'$, as
well as under the reflection $R$ and time reversal $T$. Note that, in the case
of the square lattice antiferromagnet, the displacements by one lattice spacing
induce a sign change, while under 90 degrees rotations the winding number is
invariant.

Static classical solutions minimize the energy functional given by
\begin{equation}
E[\vec e] = \int d^2x \ \frac{\rho_s}{2} \p_i \vec e \cdot \p_i \vec e.
\end{equation}
Simple vector algebra,
\begin{eqnarray}
\label{derivingSchwarz}
0&\leq&\int d^2x \ \left(\p_i \vec e \pm
\varepsilon_{ij} \p_j \vec e \times \vec e\right)^2 \nonumber \\
&=&\int d^2x \ \left(2 \p_i \vec e \cdot \p_i \vec e \pm 2
\varepsilon_{ij} \vec e \cdot \left(\p_i \vec e \times \p_j \vec e\right)\right)
= \frac{4}{\rho_s} E[\vec e] \pm 16 \pi n[\vec e],
\end{eqnarray}
leads to the Schwarz inequality
\begin{equation}
\label{schwarz}
E[\vec e] \geq 4 \pi \rho_s |n[\vec e]|.
\end{equation}
One distinguishes between Skyrmions that minimize the energy in the topological
sector with winding number $n[\vec e] = 1$, and anti-Skyrmions where
$n[\vec e] = - 1$. Classically both configurations have a rest energy of
${\cal M} c^2 = 4 \pi \rho_s$. According to Eq.~(\ref{derivingSchwarz}),
(anti-)\-Skyr\-mi\-ons satisfy the inequality Eq.~(\ref{schwarz}) as an 
equality, provided that they obey the (anti-)\-self-du\-a\-li\-ty equation
\begin{equation}
\p_i \vec e + \sigma \varepsilon_{ij} \p_j \vec e \times \vec e = 0.
\end{equation}
Note that the quantity $\sigma$ refers to either Skyrmions ($\sigma =1$) or
anti-Skyrmions ($\sigma =-1$). A particular (anti-)\-Skyr\-mi\-on 
configuration, in polar coordinates $(x_1,x_2) = r (\cos\chi,\sin\chi)$, is
\begin{equation}
\label{skyrmion}
\vec e_{\sigma,n,\rho}(r,\chi) = \left(\frac{2 r^n \rho^n}{r^{2n} + \rho^{2n}}
\cos(n \chi),\frac{2 r^n \rho^n \sigma}{r^{2n} + \rho^{2n}} \sin(n \chi),
\frac{r^{2n} - \rho^{2n}}{r^{2n} + \rho^{2n}}\right).
\end{equation}
In either case, the winding number is given by $n[\vec e] = \sigma n$, where
$n \in \N_{>0}$. The (anti-)\-Skyr\-mi\-on is centered at the origin and has 
size $\rho$.

It should be pointed out that the radial profile of the Skyrmion gets modified,
when holes localize on a Skyrmion or anti-Skyrmion. In order to take this
effect related to the details of the hole-Skyrmion interaction into account, we
will allow for a general radial profile function $f(r) \in [ -1,1 ]$ in our
ansatz for the (anti-)\-Skyr\-mi\-on configurations,
\begin{equation}
\label{skyrmionGeneral} \vec e_{\sigma,n,\rho}(r,\chi) = \left(
\sqrt{1-f^2(r)} \cos(n \chi), \sigma \sqrt{1-f^2(r)} \sin(n \chi),
f(r)\right),
\end{equation}
as we discuss in Sec.~\ref{HoleLocalization}. The behavior of the profile
function $f(r)$ at the origin and at infinity is the same as for the standard
radial profile, i.e., $f(0)=-1$ and $f(\infty)=1$.

The Skyrmion configurations defined in Eq.~(\ref{skyrmion}) are characterized
by a number of zero-modes. A shift to an arbitrary position $x$, spatial
rotations
\begin{equation}
O(\beta) x = (r \cos(\chi + \beta),r \sin(\chi + \beta),t),
\end{equation}
(where $x = (r \cos\chi,r \sin\chi,t)$) by an arbitrary angle $\beta$, or a
$U(1)_s$ spin-rotation by an arbitrary angle $\gamma$, do not alter the
Skyrmion's energy. As for the square lattice, also in the present case of the
honeycomb lattice, spatial rotations and $U(1)_s$ spin rotations acting on a
Skyrmion configuration are related by
\begin{eqnarray}
^{O(\beta)}\vec e_{\sigma,n,\rho}(r,\chi)\!\!\!\!&=&\!\!\!\!
\left(\frac{2 r^n \rho^n}
{r^{2n} + \rho^{2n}} \cos(n (\chi + \beta)),
\frac{2 r^n \rho^n \sigma}{r^{2n} + \rho^{2n}} \sin(n (\chi + \beta)),
\frac{r^{2n} - \rho^{2n}}{r^{2n} + \rho^{2n}}\right), \nonumber \\
^{I(\sigma \gamma)}\vec e_{\sigma,n,\rho}(r,\chi)\!\!\!\!&=&\!\!\!\!
\left(\frac{2 r^n \rho^n}
{r^{2n} + \rho^{2n}} \cos(n \chi + \gamma),
\frac{2 r^n \rho^n \sigma}{r^{2n} + \rho^{2n}} \sin(n \chi + \gamma),
\frac{r^{2n} - \rho^{2n}}{r^{2n} + \rho^{2n}}\right), \nonumber \\ \,
\end{eqnarray}
such that
\begin{equation}
\label{rotations}
^{I(\sigma \gamma)}\vec e_{\sigma,n,\rho}(r,\chi) = \,
^{O(\gamma/n)}\vec e_{\sigma,n,\rho}(r,\chi).
\end{equation}
Yet another zero-mode concerns dilations: under changes of the scale parameter
$\rho$, the Skyrmion energy is not altered. We can create a family of Skyrmion
configurations from the original Skyrmion defined in Eq.~(\ref{skyrmion}) by
applying a spin rotation by an angle $\sigma \gamma$ and then performing a
shift by a distance vector $x$,
\begin{equation}
\label{Skyrmionrot}
\vec e_{\sigma,n,\rho,x,\gamma}(r,\chi) = \,
^{D(x)}\left[^{I(\sigma \gamma)}\vec e_{\sigma,n,\rho}(r,\chi)\right].
\end{equation}
The transformation properties of these more general configurations under the
various unbroken symmetry transformations are
\begin{alignat}{2}
U(1)_s:&\quad &^{I(\sigma \gamma_0)}\vec e_{\sigma,n,\rho,x,\gamma}(r,\chi) &=
\vec e_{\sigma,n,\rho,x,\gamma + \gamma_0}(r,\chi), \nonumber \\
D:&\quad &^{D(x_0)}\vec e_{\sigma,n,\rho,x,\gamma}(r,\chi) &=
\vec e_{\sigma,n,\rho,x + x_0,\gamma}(r,\chi), \nonumber \\
O(\beta):&\quad &^{O(\beta)}\vec e_{\sigma,n,\rho,x,\gamma}(r,\chi) &=
\vec e_{\sigma,n,\rho,O(\beta) x,\gamma + n \beta}(r,\chi), \nonumber \\
O':&\quad &^{O'}\vec e_{\sigma,n,\rho,x,\gamma}(r,\chi) &=
\vec e_{-\sigma,n,\rho,O x,\gamma + \frac{\pi n}{3}}(r,\chi), \nonumber \\
R:&\quad &^R\vec e_{\sigma,n,\rho,x,\gamma}(r,\chi) &=
\vec e_{- \sigma,n,\rho,Rx,- \gamma}(r,\chi).
\end{alignat}
It is worth pointing out that, unlike in particle physics where Skyrmions
correspond to baryons, in antiferromagnets the Skyrmion number is not
associated with the conserved fermion number in the underlying Hubbard model.

As we have discussed in detail for the square lattice case \citep{VHJW12}, apart
from the integer winding number $n[\vec e]$, there is another topological
invariant: the Hopf number $H[\vec e]$. While the former is defined at any
instant of time, the latter describes the topology of the order parameter vector
field $\vec e(x)$ as a function of both time and space. The Hopf number
$H[\vec e] \in \Pi_3[S^2] = \Z$ takes integer values and is related to the third
homotopy group of the sphere $S^2$. The transformation properties of the Hopf
number under the various relevant symmetries are
\begin{eqnarray}
U(1)_s:&&H[^{I(\gamma)}\vec e] = H[\vec e], \nonumber \\
D_i:&&H[^{D_i}\vec e] = H[\vec e], \nonumber \\
O:&&H[^{O}\vec e] = H[\vec e], \nonumber \\
O':&&H[^{O'}\vec e] = H[\vec e], \nonumber \\
R:&&H[^R\vec e] = - H[\vec e], \nonumber \\
T:&&H[^T\vec e] = - H[\vec e].
\end{eqnarray}

The Euclidean path integral picks up an additional factor
$\exp(i \Theta H[\vec e])$ when the Hopf term is included in the dynamics. The
values of the anyon statistics angle $\Theta$ are restricted to $0$ or $\pi$
for systems which are invariant under reflection or time-reversal symmetry.
Accordingly, the Skyrmions are quantized as bosons or fermions. If reflection
and time-reversal do not represent symmetries of the systems, then $\Theta$ may
take arbitrary values, and the spin of the Skyrmions need not be integer or
half-integer. While it is expected that the Hopf term is not present in doped
cuprates \citep{Wen88,Hal88,Dom88,Fra88,Rea89}, in the present study we include
it, because we want to keep the discussion as general as possible.

In Ref.~\citep{VHJW12} we have performed the collective mode quantization of
the Skyrmion in the undoped square lattice antiferromagnet, where the standard
profile of the Skyrmion is relevant. In the present case of the honeycomb
lattice, the analysis for the undoped system is exactly the same, and the
interested reader may consult Section 2.4 of our earlier study for details.
Here we just list some essential results of that analysis.

The quantum mechanical Hamiltonian describing Skyrmions on the honeycomb
lattice takes the form
\begin{equation}
H = {\cal M} c^2 - \frac{1}{2 {\cal M}} \p_{x_i}^2 -
\frac{1}{\sqrt{2 {\cal D}(\rho)}}
\left(\p_\rho^2 + \frac{1}{\rho} \p_\rho \right)
\frac{1}{\sqrt{2 {\cal D}(\rho)}} - \frac{n^2}{2 {\cal D}(\rho) \rho^2}
\left(\p_\gamma + i n \frac{\Theta}{2 \pi}\right)^2.
\end{equation}
The explicit expressions for the rest energy, ${\cal M} c^2$, and the inertia
of the Skyrmion with respect to dilations, ${\cal D}(\rho)$, can be found in
Ref.~\citep{VHJW12}. It should be pointed out that for the standard profile,
where the quantity ${\cal D}(\rho)$ is related to the moment of inertia
${\cal I}(\rho)$ of the Skyrmion by
\begin{equation}
{\cal I}(\rho) = \frac{{\cal D}(\rho){\rho}^2}{n^2},
\end{equation}
these two quantities are logarithmically divergent in the infrared for $n=1$.
As described in Ref.~\citep{VHJW12}, one may introduce an infrared cutoff to
cure the problem.

The collective mode wave function referring to a Skyrmion or anti-Skyrmion
characterized by its winding number $\sigma n$, momentum $p_i$, and spin
$p_\gamma = \sigma m \in \Z$ amounts to
\begin{equation}
\Psi_{p,\sigma,n,m}(x,\rho,\gamma) = \exp(i p_i x_i) \exp(i \sigma m \gamma)
\psi(\rho).
\end{equation}
Including the Hopf term in our analysis, the spin operator of the Skyrmion
(i.e., the analog of isospin in particle physics) is given by
\begin{equation}
I = \sigma \left(p_\gamma + n \frac{\Theta}{2 \pi}\right) =
\sigma \left(- i \p_\gamma + n \frac{\Theta}{2 \pi}\right).
\end{equation}
The state $\Psi_{p,\sigma,n,m}(x,\rho,\gamma)$ thus corresponds to the ``isospin''
\begin{equation}
I \Psi_{p,\sigma,n,m}(x,\rho,\gamma) =
\left(m + \sigma n \frac{\Theta}{2 \pi}\right) \Psi_{p,\sigma,n,m}(x,\rho,\gamma).
\end{equation}
Note that for $\Theta = 0$ the ``isospin'' takes integer values, whereas for
$\Theta = \pi$ it is a half-integer for odd $n$.

Finally, according to Eq.~(\ref{rotations}), the angular momentum $J$ of a
Skyrmion or anti-Skyrmion turns out to be
\begin{equation}
J = \sigma n I = n \left(p_\gamma + n \frac{\Theta}{2 \pi}\right) =
n \left(- i \p_\gamma + n \frac{\Theta}{2 \pi}\right),
\end{equation}
such that
\begin{equation}
J \Psi_{p,\sigma,n,m}(x,\rho,\gamma) =
n \left(\sigma m + n \frac{\Theta}{2 \pi}\right)
\Psi_{p,\sigma,n,m}(x,\rho,\gamma).
\end{equation}
For $\Theta = 0$ the Skyrmion has integer angular momentum and hence represents
a boson, while for $\Theta = \pi$ the angular momentum takes half-odd-integer
values and the Skyrmion hence is a fermion. If the anyon statistics angle
$\Theta$ is different from 0 or $\pi$, then we are dealing with anyons in
$(2+1)$ dimensions, i.e., particles with any (neither integer nor half-integer)
angular momentum.

\section{Effective Action for Hole-Doped Antiferromagnets on the Honeycomb
Lattice}

The effective Lagrangian for hole-doped antiferromagnets on the honeycomb
lattice has been established in Ref.~\citep{Kae11}. In this section we review
some basic aspects of that systematic construction.

\subsection{Nonlinear Realization of the $SU(2)_s$ Symmetry}

One essential ingredient of the effective field theory analysis is the nonlinear
realization of the spontaneously broken $SU(2)_s$ symmetry, which allows one to
couple holes to the staggered magnetization order parameter \citep{Kae05}. The
global $SU(2)_s$ symmetry then appears as a local $U(1)_s$ symmetry in the
unbroken subgroup.

Let us parametrize the unit-vector order parameter field $\vec e_\varphi(x)$ as
\begin{equation}
\vec e_\varphi(x) = \left(- \sin \varphi(x),\cos \varphi(x),0\right).
\end{equation}
The nonlinear realization of the $SU(2)_s$ symmetry is based on the matrix
\begin{eqnarray}
\label{defu}
u(x)&=&\frac{1}{\sqrt{2 (1 + e_3(x))}}
\left(\begin{array}{cc} 1 + e_3(x) & e_1(x) - i e_2(x) \\
- e_1(x) - i e_2(x) & 1 + e_3(x) \end{array}\right) \nonumber \\
&=&\left(\begin{array}{cc} \cos\left(\frac{1}{2} \theta(x)\right) &
\sin\left(\frac{1}{2} \theta(x)\right) \exp(- i \varphi(x)) \\
- \sin\left(\frac{1}{2} \theta(x)\right) \exp(i \varphi(x)) &
\cos\left(\frac{1}{2} \theta(x)\right) \end{array}\right) \nonumber \\
&=&\cos\left(\frac{1}{2} \theta(x)\right) +
i \sin\left(\frac{1}{2} \theta(x)\right)
\vec e_\varphi(x) \cdot \vec \sigma.
\end{eqnarray}
Applying a global $SU(2)_s$ transformation $g$, the field $u(x)$ turns into
\begin{equation}
\label{trafou}
u(x)' = h(x) u(x) g^\dagger, \qquad u_{11}(x)' \geq 0.
\end{equation}
By the above equation, the nonlinear symmetry transformation,
\begin{equation}
h(x) = \exp(i \alpha(x) \sigma_3) = \left(\begin{array}{cc}
\exp(i \alpha(x)) & 0 \\ 0 & \exp(- i \alpha(x)) \end{array} \right) \in U(1)_s,
\end{equation}
is implicitly defined. We thus see that the global transformations
$g \in SU(2)_s$, related to the spontaneously broken non-Abelian spin symmetry,
manifest themselves as local transformations $h(x) \in U(1)_s$ in the unbroken
Abelian subgroup. Note that the global subgroup transformations $I(\gamma)$,
defined in Eq.~(\ref{subgroup}), take the simple form $\alpha(x) = - \gamma/2$.

To construct the nonlinear realization of the $SU(2)_s$ symmetry, one proceeds
with the diagonalizing matrix $u(x)$ defined in Eq.~(\ref{defu}). The magnon
field which then couples to the holes is the traceless anti-Her\-mi\-te\-an
field $v_\mu(x)$. It is obtained from the matrix $u(x)$ by
\begin{equation}
v_\mu(x) = u(x) \p_\mu u(x)^\dagger.
\end{equation}
Under the various symmetries, identified in the underlying microscopic $t$-$J$
model, this effective field transforms as
\begin{eqnarray}
\label{symm_magnonfields}
SU(2)_s:& \quad & v_\mu(x)' = h(x) (v_\mu(x) + {\partial}_\mu) h(x)^\dagger,
\nonumber \\
D_i:& \quad & ^{D_i}v_\mu(x) = v_\mu(x),
\nonumber\\
O:& \quad & ^Ov_1(x) =
\tau(Ox)\Big\{ \mbox{$\frac{1}{2}$} v_1(Ox)+ \mbox{$\frac{\sqrt{3}}{2}$}
v_2(Ox) +  \mbox{$\frac{1}{2}$}{\partial}_1
+ \mbox{$\frac{\sqrt{3}}{2}$}{\partial}_2 \Big\} \tau(Ox)^\dagger,
\nonumber\\
& \quad & ^Ov_2(x) =
\tau(Ox)\Big\{ - \mbox{$\frac{\sqrt{3}}{2}$}v_1(Ox)
+ \mbox{$\frac{1}{2}$}v_2(Ox)
- \mbox{$\frac{\sqrt{3}}{2}$}{\partial}_1
+ \mbox{$\frac{1}{2}$}{\partial}_2 \Big\} \tau(Ox)^\dagger,
\nonumber\\
& \quad & ^Ov_t(x) =
\tau(Ox)\Big\{  v_t(Ox) +{\partial}_t \Big\} \tau(Ox)^\dagger,
\nonumber\\
O' :& \quad & ^{O'}v_1(x) = \mbox{$\frac{1}{2}$} \Big( v_1{(Ox)}^*
+ \sqrt{3} v_2{(Ox)}^* \Big), \nonumber\\
& \quad & ^{O'}v_2(x) = \mbox{$\frac{1}{2}$} \Big( -\sqrt{3} v_1{(Ox)}^*
+ v_2{(Ox)}^* \Big), \nonumber\\
& \quad & ^{O'}v_t(x) = v_t{(Ox)}^*,\nonumber\\
R:&\quad & ^Rv_1(x) = v_1(Rx), \quad
^Rv_2(x) = -v_2(Rx), \nonumber \\
& \quad & ^Rv_t(x) = v_t(Rx),\nonumber \\
T:& \quad & ^Tv_i(x) = \tau(Tx)(v_i(Tx)+{\partial}_i)\tau(Tx)^\dagger,
\nonumber\\
& \quad & ^Tv_t(x) = -\tau(Tx)(v_t(Tx)+{\partial}_t)\tau(Tx)^\dagger.
\end{eqnarray}
Note that the matrix $\tau(x)$ is defined in Eq.~(3.20) of Ref.~\citep{Kae11}.
It is convenient to decompose $v_\mu(x)$ into an Abelian ``gauge'' field
$v_\mu^3(x)$ and two ``charged'' vector fields $v_\mu^\pm(x)$,
\begin{equation}
v_\mu(x) = i v_\mu^a(x) \sigma_a, \qquad
v_\mu^\pm(x) = v_\mu^1(x) \mp i v_\mu^2(x).
\end{equation}

Now for a Skyrmion $\vec e_{\sigma,n,\rho,0,\gamma}(r,\chi)$ centered at $x = 0$,
according to Eq.~(\ref{Skyrmionrot}), one ends up with
\begin{eqnarray}
\label{vSkyrmion}
v^3_1(r,\chi)&=&- \frac{\sigma n }{2r} (1-f(r))
\sin\chi,
\nonumber \\
v^3_2(r,\chi)&=&\frac{\sigma n }{2 r}(1-f(r)) \cos\chi,
\nonumber \\
v^3_t(r,\chi)&=&\frac{\sigma }{2}(1-f(r))\dot \gamma,
\nonumber \\
v^\pm_1(r,\chi)&=& \frac 1 2
\left(\mp \frac{i f'(r)}{\sqrt{1-f^2(r)}}\cos\chi - \frac{\sigma
n}{r}\sqrt{1-f^2(r)}\sin\chi\right)
\exp(\mp i \sigma \left[n \chi + \gamma\right]), \nonumber \\
v^\pm_2(r,\chi)&=&\frac 1 2
\left(\mp \frac{i f'(r)}{\sqrt{1-f^2(r)}}\sin\chi + \frac{\sigma
n}{r}\sqrt{1-f^2(r)}\cos\chi\right)
\exp(\mp i \sigma \left[n \chi + \gamma\right]), \nonumber \\
v^\pm_t(r,\chi)&=&\frac{\sigma}{2} \sqrt{1-f^2(r)} \exp(\mp i
\sigma (n \chi + \gamma)) \dot \gamma.
\end{eqnarray}
Remember that the quantity $f(r)$, introduced in Eq.~(\ref{skyrmionGeneral}),
describes a general radial profile function of the Skyrmion. It has to be
pointed out that holes, when they get localized on a Skyrmion, do affect its
radial profile. In the present study, we mainly focus on symmetry
considerations, which are not affected by the actual form of the profile $f(r)$.

\subsection{Hole Fields and Transformation Properties}

In the effective field theory description, the holes are represented by
Grassmann-valued fields $\psi^f_\pm(x)$ \citep{Bru06}. The index
$f \in \{\alpha,\beta\}$ denotes the flavor of the momentum space pockets in
which the holes reside. The subscript $\pm$, on the other hand, refers to the
spin of the hole with respect to the direction of the local staggered
magnetization. The transformation properties of the hole fields under the
symmetries of the underlying honeycomb lattice antiferromagnet are as follows:
\begin{alignat}{2}
\label{finalholefieldtrafos}
SU(2)_s:&\quad &\psi^{f}_\pm(x)' &= \exp(\pm i \alpha(x)) \psi^{f}_\pm(x),
\nonumber \\
U(1)_Q:&\quad &^Q\psi^{f}_\pm(x) &= \exp(i \omega) \psi^{f}_\pm(x),
\nonumber \\
D_i:&\quad &^{D_i}\psi^{f}_\pm(x) &= \exp(ik^{f}a_{i})\psi^{f}_\pm(x), \nonumber \\
O:&\quad &^O\psi^{\alpha}_\pm(x) &= \mp \exp(\pm i\tfrac{2\pi}{3}\mp i\varphi(Ox))\psi^{\beta}_{\mp}(Ox), \nonumber \\
&\quad &^O\psi^{\beta}_\pm(x)&=\mp \exp(\mp i\tfrac{2\pi}{3}\mp i \varphi(Ox))\psi^{\alpha}_{\mp}(Ox), \nonumber \\
O':&\quad &^{O'}\psi^{\alpha}_\pm(x) &= \pm \exp(\pm i\tfrac{2\pi}{3})\psi^{\beta}_{\mp}(Ox), \nonumber \\
&\quad &^{O'}\psi^{\beta}_\pm(x)&=\pm \exp(\mp i\tfrac{2\pi}{3})\psi^{\alpha}_\mp(Ox), \nonumber \\
R:&\quad &^R\psi^{f}_\pm(x) &= \psi^{f'}_\pm(Rx).
\end{alignat}

\begin{figure}[t]
\begin{center}
\vspace{-0.15cm}
\epsfig{file=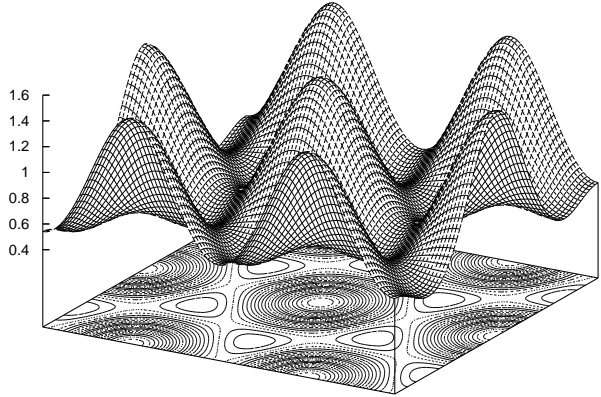,angle=-90,width=10cm} \vskip-0.5cm
\end{center}
\caption{\it Dispersion relation $E(k)/t$ of a single hole doped into a
honeycomb-lattice antiferromagnet, obtained from a simulation of the $t$-$J$
model for $J/t=2$ \citep{Jia08}.}
\label{landscapeHoneycomb}
\end{figure}

While the $U(1)_Q$ symmetry refers to fermion number conservation, the
quantities $k^\alpha = - k^\beta = \mbox{$(0,\frac{4\pi}{3\sqrt{3} a})$}$
represent the lattice momenta of the pockets where doped holes of flavor
$\alpha$ and $\beta$ reside. As illustrated in Fig.~\ref{landscapeHoneycomb},
holes indeed occupy spherically shaped hole pockets characterized by lattice
momenta $\mbox{$(\pm\frac{2\pi}{3a}, \pm\frac{2\pi}{3\sqrt{3} a})$}$ and
$\mbox{$(0,\pm \frac{4\pi}{3\sqrt{3} a})$}$.

\subsection{Effective Action}

Using the nonlinear realization of the magnon fields and the effective
Grassmann field representation for the holes, it is straightforward to write
down the leading and subleading terms of the effective action for a hole-doped
antiferromagnet on the honeycomb lattice \citep{Kae11}. As in our earlier study
of hole localization on a Skyrmion in a square-lattice antiferromagnet, here we
also restrict ourselves to the leading terms in the effective action. Note
however, that on the square lattice hole pockets are elliptically shaped. In
Ref.~\citep{Kae11} we considered the idealized case of circular hole pockets, in
order to be able to perform large parts of the calculation analytically. Here,
on the honeycomb lattice, hole pockets have a circular shape and we do not have
to make any idealizations.

The action for hole-doped antiferromagnets on the honeycomb lattice then takes
the form
\begin{eqnarray}
\label{holeaction}
S\left[\psi^{f\dagger}_\pm,\psi^f_\pm,{\vec e}\right] & = & \int d^2x \ dt
\Bigg\{ \frac{\rho_s}{2}
\Big( \partial_i {\vec e} \cdot \partial_i {\vec e} + \frac{1}{c^2} \partial_t
{\vec e} \cdot \partial_t {\vec e}\Big)
+ \sum_{\ontopof{f=\alpha,\beta}{\, s = +,-}} \Big[ M \psi^{f\dagger}_s \psi^f_s
\nonumber \\
& + & \psi^{f\dagger}_s D_t \psi^f_s +\frac{1}{2 M'} D_i \,
\psi^{f\dagger}_s D_i \psi^f_s
+ \Lambda \psi^{f\dagger}_s (i s v^s_1 + \sigma_f v^s_2) \psi^f_{-s} \Big] \Bigg\}.
\end{eqnarray}
The quantities $M$ and $M'$ are the rest mass and the kinetic mass of a hole,
respectively. The low-energy effective coupling constant $\Lambda$ represents a
hole-one-magnon coupling which, along with all other low-energy constants,
takes real values. The sign $\sigma_f$ is $+$ for $f=\alpha$ and $-$ for
$f=\beta$. The covariant derivatives are
\begin{align}
\label{kovardrhole}
D_t \psi^f_\pm(x) & = \left[\p_t \pm i v_t^3(x) - \mu \right] \psi^f_\pm(x),
\nonumber \\
D_i \psi^f_\pm(x) &  = \left[\p_i \pm i v_i^3(x)\right] \psi^f_\pm(x).
\end{align}
Note that the chemical potential $\mu$ appears as an imaginary constant vector
potential for fermion number $U(1)_Q$ in the covariant time derivative.

Remarkably, the $\Lambda$-term which contains just a single (uncontracted)
spatial derivative is invariant under all the symmetries listed in
Eq.~(\ref{finalholefieldtrafos}). As it involves just one derivative, it
represents the leading contribution to the low-energy dynamics of a lightly
hole-doped honeycomb antiferromagnet. Note that for holed-doped square lattice
antiferromagnets, such a term --- the so-called Shraiman-Siggia term --- is
also present \citep{Bru06}, whereas it is not allowed by symmetry reasons in the
case of electron-doped square lattice antiferromagnets \citep{Bru07a}.

\subsection{Implications of the Shraiman-Siggia Term}
\label{shraimanSiggia}

The Shraiman-Siggia term dominates the dynamics of the system at low energies.
The explicit structure of this term depends both on the lattice geometry and on
the localization of the hole pockets \citep{Bru06,Kae11}. In the case of the
honeycomb lattice it takes the form
$\Lambda \psi^{f\dagger}_s (i s v^s_1 + \sigma_f v^s_2) \psi^f_{-s}$, while for the
square lattice it is given by
$\Lambda \psi^{f\dagger}_s ( v^s_1 + \sigma_f v^s_2) \psi^f_{-s}$.
As pointed out in Ref.~\citep{Kae11}, the $\Lambda$-term on the honeycomb
lattice is invariant under an accidental global rotation symmetry by an
arbitrary angle $\gamma$. For the square lattice, on the other hand, the
$\Lambda$-term is only invariant under discrete rotations by 90 degrees.

This accidental rotation symmetry emerging on the honeycomb lattice has several
important implications. First, the combination of magnon ``gauge'' fields in
the $\Lambda$-term, according to Eqs.~(\ref{vSkyrmion}), implies
\begin{eqnarray}
\label{LambdaTermImplications} i v_1^+(x) + \sigma_f v_2^+(x) & =
& \frac{1}{2} \Big[ \frac{f'(r)}{\sqrt{1-f^2(r)}} + \frac{\sigma
\sigma_f n}{r} \sqrt{1-f(r)^2} \Big]
\nonumber \\
& & \times \exp\Big(- i \sigma (n \chi + \gamma) - i \sigma_f \chi
\Big),
\nonumber \\
-i v_1^-(x) + \sigma_f v_2^-(x) & = & \frac{1}{2} \Big[
\frac{f'(r)}{\sqrt{1-f^2(r)}} + \frac{\sigma \sigma_f n}{r}
\sqrt{1-f(r)^2} \Big]
\nonumber \\
& & \times
\exp\Big(i \sigma (n \chi + \gamma) + i \sigma_f \chi \Big),
\end{eqnarray}
and leads to the following interesting effect. The (anti-)\-Skyr\-mi\-on 
standard profile, defined in Eq.~(\ref{skyrmion}), satisfies the differential 
equation
\begin{equation}
\frac{f'(r)}{\sqrt{1-f^2(r)}} - \frac{n}{r} \sqrt{1-f^2(r)} = 0,
\qquad f(r)=\frac{r^{2n} - \rho^{2n}}{r^{2n} + \rho^{2n}}.
\end{equation}
Accordingly, for the standard profile, a Skyrmion can only localize a hole with
flavor $\alpha$, while an anti-Skyrmion can only localize a hole of flavor
$\beta$. For a general profile $f(r)$ on the honeycomb lattice, there is no
such restriction. Note that on the square lattice, holes of flavor $\alpha$ and
$\beta $ can localize both on Skyrmions and anti-Skyrmions for a general 
Skyrmion profile, including the standard one.

Another advantage of the honeycomb geometry is that the wave function of a
single hole or of two holes localized on a (anti-)\-Skyr\-mi\-on, factorizes 
into a radial and an angular part for an arbitrary profile function $f(r)$. As 
we will see in the next section, large parts of our calculations can thus be 
performed analytically. In the case of the square lattice, where an accidental 
rotation symmetry is absent, the factorization only occurs for the standard 
profile.

\section{Hole Localization on a Skyrmion}
\label{HoleLocalization}

This section deals with the application of the effective theory established in
the previous section to the localization of holes on a Skyrmion. First, we
consider the localization of a single hole, both on a static and on a rotating
Skyrmion. Afterwards we investigate the localization of two holes of the same
flavor on the same (anti-)\-Skyr\-mi\-on, and analyze the symmetry properties 
of the emerging two-hole bound states.

\subsection{Single Hole Localized on a Static Skyrmion}

As it was shown in \citep{VHJW12}, the Skyrmion's moment of inertia
${\cal I}(\rho)$ is logarithmically divergent in the infrared for $n=1$ and for
the standard profile. In this case, unless the divergence is regularized, the
Skyrmion cannot rotate. Since in this study we focus on symmetry aspects and
not on the details of the Skyrmion dynamics, we neglect the Skyrmion's
translational and dilational motion, and fix the center of the Skyrmion at the
origin $x = 0$. We also fix the size of the Skyrmion to a constant radius
$\rho$. If holes are incorporated, the energy of the Skyrmion-hole bound states
takes a minimum for a particular value of $\rho$, as we will see later on.

If a single hole is localized on a Skyrmion, the corresponding wave function
amounts to
\begin{equation}
\Psi^f_{\sigma,n}(r,\chi) = \left(\begin{array}{c}
\Psi^f_{\sigma,n,+}(r,\chi) \\ \Psi^f_{\sigma,n,-}(r,\chi)
\end{array} \right).
\end{equation}
Since the rest energy $M$ of the holes just corresponds to a constant energy
shift, this term can be neglected. The Hamiltonian resulting from the action of
Eq.(\ref{holeaction}) thus takes the form
\begin{eqnarray}
\label{hamiltonian0}
H^f&=&\left(\begin{array}{cc} H^f_{++} & H^f_{+-} \\
H^f_{-+} & H^f_{--} \end{array} \right), \nonumber \\
H^f_{++}&=&- \frac{1}{2 M'} \left[\p_i + i v^3_i(x)\right]^2 = -
\frac{1}{2 M'} \left[\p_r^2 + \frac{1}{r} \p_r + \frac{1}{r^2}
\left( \p_\chi + i \frac{\sigma n}{2} (1-f(r))
\right)^2 \right], \nonumber \\
H^f_{+-}&=& \Lambda(i v_1^+(x) + \sigma_f v_2^+(x)) \nonumber \\
&=&\frac{\Lambda}{2} \Big[ \frac{f'(r)}{\sqrt{1-f^2(r)}} +
\frac{\sigma \sigma_f n}{r} \sqrt{1-f^2(r)} \Big] \exp\Big(- i
\sigma \Big[(n +\sigma \sigma_f) \chi + \gamma  \Big] \Big),
\nonumber \\
H^f_{-+}&=&\Lambda (-i v_1^-(x) + \sigma_f v_2^-(x)) \nonumber \\
&=&\frac{\Lambda}{2} \Big[ \frac{f'(r)}{\sqrt{1-f^2(r)}} +
\frac{\sigma \sigma_f n}{r} \sqrt{1-f^2(r)} \Big] \exp\Big( i
\sigma \Big[(n +\sigma \sigma_f) \chi + \gamma  \Big] \Big),
\nonumber \\
H^f_{--}&=& - \frac{1}{2 M'} \left[\p_i - i v^3_i(x)\right]^2 = -
\frac{1}{2 M'} \left[\p_r^2 + \frac{1}{r} \p_r + \frac{1}{r^2}
\left(\p_\chi - i \frac{\sigma n}{2} (1-f(r)) \right)^2 \right].
\qquad
\end{eqnarray}
With the explicit expressions for $v_i^3(x)$ and $v_i^\pm(x)$ of the Skyrmion
provided in Eq.(\ref{vSkyrmion}), and making the ansatz
\begin{equation}
\label{ansatz1} \Psi^f_{\sigma,m_+^f,m_-^f}(r,\chi) =
\left(\begin{array}{c} \psi^f_{\sigma,m_+^f,m_-^f,+}(r)
\exp\left(i \sigma [m_+^f \chi - \frac{\gamma}{2}] \right) \\
\psi^f_{\sigma,m_+^f,m_-^f,-}(r) \exp\left(i \sigma [m_-^f \chi +
\frac{\gamma}{2}] \right) \end{array} \right),
\end{equation}
with $m_-^f - m_+^f = n + \sigma \sigma_f$, one readily derives the radial
Schr\"odinger equation
\begin{equation}
H_r^f \psi^f_{\sigma,m_+^f,m_-^f}(r) = \left(\begin{array}{cc} H_{r++}^f &
H_{r+-}^f \\ H_{r-+}^f & H_{r--}^f \end{array} \right)
\left(\begin{array}{c} \psi^f_{\sigma,m_+^f,m_-^f,+}(r) \\
\psi^f_{\sigma,m_+^f,m_-^f,-}(r)
\end{array} \right) = E_{m_+^f,m_-^f} \psi^f_{\sigma,m_+^f,m_-^f}(r),
\end{equation}
with
\begin{eqnarray}
&&H_{r++}^f = - \frac{1}{2 M'} \left[\p_r^2 + \frac{1}{r} \p_r -
\frac{1}{r^2} \left(m_+^f + \frac{n}{2} (1-f(r))
\right)^2\right],
\nonumber \\
&&H_{r+-}^f = H_{r-+}^f =
\frac{\Lambda}{2} \Big[ \frac{f'(r)}{\sqrt{1-f^2(r)}} + \frac{\sigma
\sigma_f n}{r} \sqrt{1-f^2(r)} \Big], \nonumber \\
&&H_{r--}^f =  - \frac{1}{2 M'} \left[\p_r^2 + \frac{1}{r} \p_r -
\frac{1}{r^2} \left(m_-^f - \frac{n}{2} (1-f(r))
\right)^2\right].
\end{eqnarray}
It should be pointed out that the emerging radial Schr\"odinger equation is not
the same for Skyrmions and anti-Skyrmions, and neither is it identical for
flavors $f = \alpha,\beta$. In the case of the square lattice, on the other
hand, the resulting radial Schr\"odinger equation is the same for both flavors
as well as for Skyrmions and anti-Skyrmions \citep{VHJW12}. Interestingly, if
$n$ is odd and if $m_-^f = - m_+^f = (n+\sigma \sigma_f)/2$, the two equations
decouple, and the equation that describes a localized hole amounts to
\begin{eqnarray}
\label{decoupled_equation} \Bigg\{- \frac{1}{2 M'}\left(\p_r^2 +
\frac{1}{r} \p_r - \frac{1}{r^2} \left(\frac{n + \sigma
\sigma_f}{2} - \frac{n }{2}(1-f(r))\right)^2 \right)-\nonumber \\
 \frac{\Lambda}{2}\left(\frac{f'(r)}{\sqrt
{1-f^2(r)}}+\frac{\sigma \sigma_f n}{r}\sqrt{1-f^2(r)}\right)
\Bigg\}\psi^{f}(r)= E \psi^{f}(r),
\end{eqnarray}
where $\psi^f(r)$ corresponds to the linear combination
\begin{equation}
\label{decoupled_wave} \psi^f(r) =
\frac{1}{\sqrt{2}}\left(\psi^f_{\sigma,m_+^f,m_-^f,+}(r) -
\psi^f_{\sigma,m_+^f,m_-^f,-}(r)\right).
\end{equation}
In the present study, we will be most interested in (anti-)\-Skyr\-mi\-ons with
winding number $\sigma n = \pm 1$. For a Skyrmion or an anti-Skyrmion with
$n=1$, the two equations decouple, but are still different for different 
flavors. For the Skyrmion (with $n=1$ and $\sigma = 1$) the radial 
Schr\"odinger equation amounts to
\begin{equation}
\left[-\frac{1}{2 M'} \left(\p_r^2 + \frac{1}{r} \p_r\right) +
V^f(r)\right] \psi^f(r) = E \psi^f(r),
\end{equation}
\begin{equation}
\label{potential alpha} V^\alpha(r)=\frac{1}{8 M'r^2} (1+f(r))^2
 - \frac{\Lambda}{2}\left(\frac{f'(r)}{\sqrt
{1-f^2(r)}}+\frac{1}{r}\sqrt{1-f^2(r)}\right),
\end{equation}
\begin{equation}
\label{potential beta} V^\beta(r)= \frac{1}{8 M'r^2} (1-f(r))^2
 - \frac{\Lambda}{2}\left(\frac{f'(r)}{\sqrt
{1-f^2(r)}}-\frac{1}{r}\sqrt{1-f^2(r)}\right).
\end{equation}
For the standard radial profile of the Skyrmion, given by
$f(r)={(r^{2n}-\rho^{2n})}/{(r^{2n}+\rho^{2n})}$, only $\alpha$-holes can be
localized on the Skyrmion, since $\beta$-holes have a repulsive potential.
Vice versa, $\alpha$-holes can not be localized on the anti-Skyrmion, but
$\beta$-holes can be attracted by an anti-Skyrmion.

\begin{figure}[t]
\begin{center}
\epsfig{file=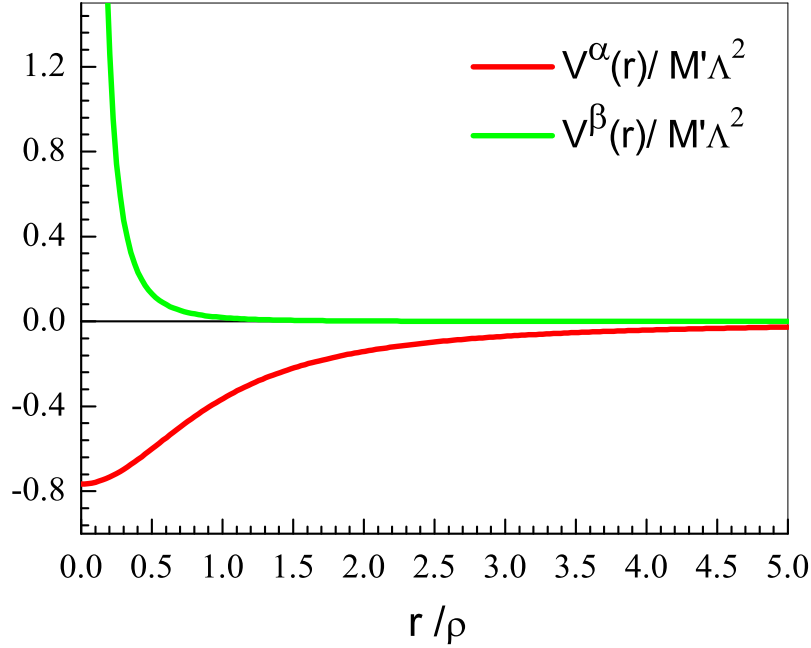,width=12cm} \vspace{-1.0cm}
\end{center}
\caption{\it (Color online) The potentials $V^\alpha(r)$ and $V^\beta(r)$ related
to the standard Skyrmion profile for $n=1$.}
\label{standardProfilePlot}
\end{figure}

Although the main focus of the present study is a careful symmetry analysis, we 
still want to get a rough idea on the energy scales involved. Let us consider 
the standard Skyrmion profile and the situation where only one hole is 
localized. Here an attractive potential only emerges for an $\alpha$-hole 
localized on a Skyrmion (or a $\beta$-hole localized on an anti-Skyrmion), 
while the potential is repulsive in the other channel. Both the attractive and 
the repulsive potential are illustrated in Fig.~\ref{standardProfilePlot}.
For $n = 1$, the
radial Schr\"odinger equation reduces to
\begin{equation}
\left[- \frac{1}{2 M'} \left(\p_r^2 + \frac{1}{r} \p_r\right) + V^{\alpha}(r)
\right] \psi^{\alpha}(r) = E \psi^{\alpha}(r),
\end{equation}
with the potential
\begin{equation}
V^{\alpha}(r) = \frac{1}{2 M'} \frac{r^2}{(r^2 + \rho^2)^2} -
2 \Lambda \frac{\rho}{r^2 + \rho^2}.
\end{equation}
We now use the harmonic oscillator approximation, where at short distances the
potential takes the form
\begin{equation}
V^{\alpha}_{\text{approx}}(r) = - \frac{2 \Lambda}{\rho} + \frac{M'}{2}
\left(\frac{1}{{M'}^2 \rho^4} + \frac{4 \Lambda}{M' \rho^3}\right) r^2
+ {\cal O}(r^4).
\end{equation}
In this rather crude approximation, the ground state energy amounts to
\begin{equation}
E_0 = - \frac{2\Lambda}{\rho} +
\sqrt{\frac{1}{{M'}^2 \rho^4} + \frac{4 \Lambda}{M' \rho^3}} =
M' \Lambda^2 x \left(\sqrt{x^2 + 4 x} - 2\right), \quad
x = \frac{1}{M' \Lambda \rho}.
\end{equation}
In terms of the parameter $x$, the minimization of the energy yields
\begin{eqnarray}
\label{rho}
&&x = \frac{2}{\sqrt{3}} \left[
\left(\frac{3 \sqrt{3}}{4} + \frac{\sqrt{11}}{4}\right)^{1/3} +
\left(\frac{3 \sqrt{3}}{4} + \frac{\sqrt{11}}{4}\right)^{-1/3}\right] - 2
\approx 0.383 \ \Rightarrow \nonumber \\
&&\rho \approx \frac{1}{0.383 M' \Lambda}.
\end{eqnarray}
The emerging bound state with the strongest binding energy is characterized by
\begin{equation}
\label{bindingE}
E_0 = M' \Lambda^2 x \left(\sqrt{x^2 + 4 x} - 2\right) \approx
- 0.270 M' \Lambda^2.
\end{equation}
\begin{figure}[t]
\begin{center}
\epsfig{file=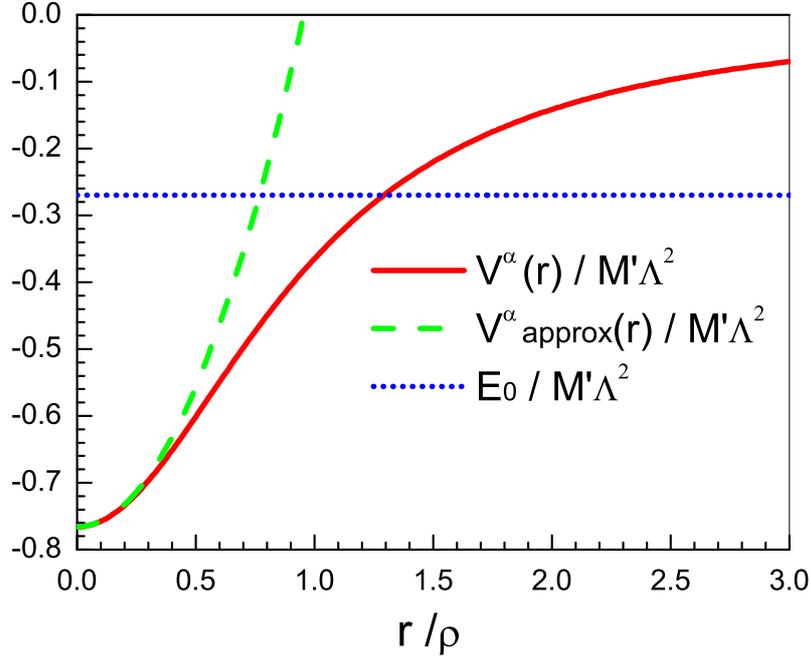,width=12cm} \vspace{-1.0cm}
\end{center}
\caption{\it (Color online) The potential $V^{\alpha}(r)$ (solid line) along
with its harmonic approximation (dashed line) and the resulting ground-state
energy (dotted line).}
\label{alphaPotentialPlot}
\end{figure}
In Fig.~\ref{alphaPotentialPlot} we have plotted the potential $V^{\alpha}(r)$
along with its harmonic approximation and the ground state energy $E_0$.
According to this figure, the true energy of the ground state seems to be
smaller than the harmonic approximation suggests.

\subsection{Single Hole Localized on a Rotating Skyrmion}

The Lagrange function for the rotational motion that involves $\gamma$ is given
by
\begin{equation}
L = \frac{{\cal D}(\rho) \rho^2}{2 n^2}\dot{\gamma}^2 -
n \frac{\Theta}{2 \pi} \dot \gamma +
\int d^{2}x \sum_{\ontopof{f=\alpha,\beta}{\, s = +,-}}
s \psi^{f \dagger}_s v^3_t \psi^f_s.
\end{equation}
The momentum canonically conjugate to $\gamma$, using Eq.(\ref{vSkyrmion}),
hence amounts to
\begin{equation}
p_\gamma = \frac{{\cal D}(\rho) \rho^2 \dot \gamma}{n^2} - n
\frac{\Theta}{2 \pi} + \int d^{2}x \ \frac{\sigma}{2}(1-f(r))
\sum_{\ontopof{f=\alpha,\beta}{\, s = +,-}} s \psi^{f \dagger}_s
\psi^f_s,
\end{equation}
leading to the Hamiltonian
\begin{equation}
H^\gamma = \frac{1}{2 {\cal I}(\rho)}
\left(- i \partial_\gamma - A_\gamma\right)^2,
\end{equation}
where $A_\gamma$ is the Berry gauge field given by
\begin{equation}
A_\gamma = \int d^2x \sum_{\ontopof{f=\alpha,\beta}{s = +,-}}
\Psi^{f \dagger}_s \frac{\sigma}{2}(1-f(r)) s \Psi^f_s - n
\frac{\Theta}{2 \pi}.
\end{equation}
Combining the above results, one notices that the off-diagonal elements of the
Hamiltonian (\ref{hamiltonian0}) remain the same, whereas the diagonal elements
involve additional contributions:
\begin{eqnarray}
\label{hamiltonian}
H^f&=&\left(\begin{array}{cc} H^f_{++} & H^f_{+-} \\
H^f_{-+} & H^f_{--} \end{array} \right), \nonumber \\
H^f_{++}&=&- \frac{1}{2 M'} \left[\p_r^2 + \frac{1}{r} \p_r +
\frac{1}{r^2} \left( \p_\chi + i \frac{\sigma n}{2} (1-f(r))
\right)^2 \right]\nonumber \\
&-&\frac{n^2}{2 {\cal D}(\rho) \rho^2} \left(\p_\gamma + i n
\frac{\Theta}{2 \pi} - i
\frac{\sigma}{2}(1-f(r))\right)^2, \nonumber \\
H^f_{+-}&=& \frac{\Lambda}{2} \Big[ \frac{f'(r)}{\sqrt{1-f^2(r)}}
+ \frac{\sigma \sigma_f n}{r} \sqrt{1-f^2(r)} \Big] \exp\Big(- i
\sigma \Big[(n +\sigma \sigma_f) \chi + \gamma  \Big] \Big),
\nonumber \\
H^f_{-+}&=&\frac{\Lambda}{2} \Big[ \frac{f'(r)}{\sqrt{1-f^2(r)}} +
\frac{\sigma \sigma_f n}{r} \sqrt{1-f^2(r)} \Big] \exp\Big( i
\sigma \Big[(n +\sigma \sigma_f) \chi + \gamma  \Big] \Big),
\nonumber \\
H^f_{--}&=&- \frac{1}{2 M'} \left[\p_r^2 + \frac{1}{r} \p_r +
\frac{1}{r^2} \left( \p_\chi - i \frac{\sigma n}{2} (1-f(r))
\right)^2 \right]\nonumber \\
&-&\frac{n^2}{2 {\cal D}(\rho) \rho^2} \left(\p_\gamma + i n
\frac{\Theta}{2 \pi} + i \frac{\sigma}{2}(1-f(r))\right)^2.
\nonumber \\ \qquad
\end{eqnarray}
Our ansatz for the wave function is
\begin{equation} \label{ansatz2}
\Psi^f_{\sigma,m_+^f,m_-^f,m}(r,\chi,\gamma)\! =\!
\left(\begin{array}{c} \psi^f_{\sigma, m_+^f,m_-^f,m,+}(r)
\exp(i \sigma m_+^f \chi) \exp\left(i \sigma (m - \frac{1}{2}) \gamma\right) \\
\psi^f_{\sigma, m_+^f,m_-^f,m,-}(r) \exp(i \sigma m_-^f \chi)
\exp\left(i \sigma (m + \frac{1}{2}) \gamma\right) \\
\end{array} \right),
\end{equation}
with $m_-^f - m_+^f = n + \sigma\sigma_f$. Since the wave function must be
$2 \pi$-periodic in the variable $\gamma$, the quantum number $m$ must now be
one half of some odd integer. Note that in the case of the rotating Skyrmion
without a hole, discussed in Subsection 2.2 (collective mode quantization),
$m$ was an integer. The radial Schr\"odinger equation then takes the form
\begin{eqnarray}
H_r^f \psi^f_{\sigma,m_+^f,m_-^f,m}(r)&=&\left(\begin{array}{cc}
H_{r++}^f & H_{r+-}^f \\ H_{r-+}^f & H_{r--}^f \end{array} \right)
\left(\begin{array}{c} \psi^f_{\sigma,m_+^f,m_-^f,m,+}(r) \\
\psi^f_{\sigma,m_+^f,m_-^f,m,-}(r)
\end{array} \right) \nonumber \\
&=&E_{\sigma,m_+^f,m_-^f,m} \psi^f_{\sigma,m_+^f,m_-^f,m}(r),
\end{eqnarray}
where the four matrix elements of the radial Hamiltonian $H^f_r$ are
\begin{eqnarray}
\label{radialeq}
H_{r++}^f&=&- \frac{1}{2 M'} \left[\p_r^2 +
\frac{1}{r} \p_r - \frac{1}{r^2} \left(m_+^f + \frac{n}{2}
(1-f(r))\right)^2\right]
\nonumber \\
&+&\frac{n^2}{2 {\cal D}(\rho) \rho^2} \left(m - \frac{1}{2} + \sigma n
\frac{\Theta}{2 \pi} - \frac{1}{2}(1-f(r))
\right)^2, \nonumber \\
H_{r+-}^f&=&H_{r-+}^f = \frac{\Lambda}{2} \Big[
\frac{f'(r)}{\sqrt{1-f^2(r)}}
+ \frac{\sigma \sigma_f n}{r} \sqrt{1-f^2(r)} \Big] , \nonumber \\
H_{r--}^f&=&- \frac{1}{2 M'} \left[\p_r^2 + \frac{1}{r} \p_r -
\frac{1}{r^2} \left(m_-^f - \frac{n}{2} (1-f(r))\right)^2\right]
\nonumber \\
&+&\frac{n^2}{2 {\cal D}(\rho) \rho^2} \left(m + \frac{1}{2} + \sigma n
\frac{\Theta}{2 \pi} + \frac{1}{2}(1-f(r))\right)^2.
\end{eqnarray}

\subsection{Single Hole Localized on a Rotating Skyrmion: Symmetry
Properties}

Recall that the spin operator (which is related to an internal symmetry
transformation and hence is analogous to isospin in particle physics)
\begin{equation}
I = \left(\begin{array}{cc} - i \sigma \p_\gamma + \sigma n \frac{\Theta}{2 \pi}
+ \frac{1}{2} & 0 \\ 0 & - i \sigma \p_\gamma + \sigma n \frac{\Theta}{2 \pi} -
\frac{1}{2} \end{array}\right),
\end{equation}
commutes with the Hamiltonian: $[H^f,I] = 0$. The wave function
$\Psi^f_{\sigma,m^f_+,m^f_-,m}$ is then an eigenstate of $I$, i.e.
\begin{equation}
I \ \Psi^f_{\sigma,m^f_+,m^f_-,m}(r,\chi,\gamma) =
\left(m + \sigma n \frac{\Theta}{2 \pi}\right)
\Psi^f_{\sigma,m^f_+,m^f_-,m}(r,\chi,\gamma).
\end{equation}
Because $m$ is half of an odd integer, the rotating Skyrmion with one hole
localized on it carries half-integer spin (or ``isospin''), provided that the
anyon statistics parameter vanishes: $\Theta = 0$.

Under the various discrete symmetries --- the displacements $D_1$ and $D_2$, the
rotation $O'$, and the reflection $R$ --- the wave function
\begin{equation}
\Psi^f_{\sigma,n}(r,\chi,\gamma) = \left(\begin{array}{c}
\Psi^f_{\sigma,n,+}(r,\chi,\gamma) \\ \Psi^f_{\sigma,n,-}(r,\chi,\gamma)
\end{array} \right),
\end{equation}
describing a single hole localized on a rotating (anti-)\-Skyr\-mi\-on with 
winding number $\sigma n$, transforms as
\begin{eqnarray}
&&^{D_i}\Psi^f_{\sigma,n}(r,\chi,\gamma) = \exp(i k^f a_i)
\left(\begin{array}{c} \Psi^f_{\sigma,n,+}(r,\chi,\gamma) \\
\Psi^f_{\sigma,n,-}(r,\chi,\gamma)
\end{array} \right), \nonumber \\
&&^{O'}\Psi^f_{\sigma,n}(r,\chi,\gamma) = \left(\begin{array}{c}
\exp(-\sigma_f \frac{2\pi i}{3})\Psi^f_{\sigma,n,-}(r,\chi + \frac{\pi}{3},\gamma
- n \frac{\pi}{3}) \\
-\exp(\sigma_f \frac
{2\pi i}{3})\Psi^f_{\sigma,n,+}(r,\chi +
\frac{\pi}{3},\gamma - n \frac{\pi}{3})
\end{array} \right), \nonumber \\
&&^R\Psi^f_{\sigma,n}(r,\chi,\gamma) =
\left(\begin{array}{c}
\Psi^f_{\sigma,n,+}(r,- \chi,- \gamma) \\
\Psi^f_{\sigma,n,-}(r,- \chi,- \gamma) \end{array} \right).
\end{eqnarray}
Accordingly, the energy eigenstates transform as
\begin{eqnarray}
&&^{D_i}\Psi^f_{\sigma,m_+^f,m_-^f,m}(r,\chi,\gamma) =  \exp(i k^f
a_i)
\Psi^f_{\sigma,m_+^f,m_-^f,m}(r,\chi,\gamma), \nonumber \\
&&^{O'}\Psi^\alpha_{\sigma,m_+^\alpha,m_-^\alpha,m}(r,\chi,\gamma) =
\exp\left(i \sigma [m_+^\alpha + \frac n 2-\sigma-mn]\frac \pi 3
\right)
\Psi^\beta_{-\sigma,-m_-^\alpha,-m_+^\alpha,-m}(r,\chi,\gamma), \nonumber \\
&&^{O'}\Psi^\beta_{\sigma,m_+^\beta,m_-^\beta,m}(r,\chi,\gamma) =
\exp\left(i \sigma [m_+^\beta + \frac n 2+\sigma-mn]\frac \pi 3
\right)
\Psi^\alpha_{-\sigma,-m_-^\beta,-m_+^\beta,-m}(r,\chi,\gamma), \nonumber \\
&&^R\Psi^\alpha_{\sigma,m_+^\alpha,m_-^\alpha,m}(r,\chi,\gamma) =
\Psi^\beta_{-\sigma,m_+^\alpha,m_-^\alpha,m}(r,\chi,\gamma), \nonumber \\
&&^R\Psi^\beta_{\sigma,m_+^\beta,m_-^\beta ,m}(r,\chi,\gamma) =
\Psi^\alpha_{-\sigma,m_+^\beta,m_-^\beta,m}(r,\chi,\gamma).
\end{eqnarray}
Note that for $\Theta \neq 0$ or $\pi$, the Hopf term explicitly breaks the
reflection symmetry $R$. Deriving these equations, we have used appropriate
phase conventions for the radial wave functions. For the rotation symmetry
$O'$, we have used $\alpha'=\beta$, $\beta'=\alpha$, as well as
\begin{equation}
\psi^f_{\sigma,m_+^f,m_-^f,m,\mp}(r)=\psi^{f'}_{-\sigma,-m_-^{f},-m_+^{f},-m,\pm}(r),
\end{equation}
which results from the behavior of Eq.(\ref{radialeq}) by making the
replacements $m_{+}^{f} \rightarrow m_{+}^{f'} = - m_{-}^{f}$,
$m_{-}^{f} \rightarrow m_{-}^{f'} = - m_{+}^{f}$, and $m \rightarrow m' = - m$.
For the reflection $R$ we have used
\begin{equation}
\psi^f_{\sigma,m_+^f,m_-^f,m,\pm}(r)=\psi^{f'}_{-\sigma,m_+^{f},m_-^{f},m,\pm}(r).
\end{equation}
Note that after these replacements the constraint
\begin{equation}
m_{-}^{f} - m_{+}^{f} = n + \sigma\sigma_f
\end{equation}
is still satisfied.

\subsection{Schr\"odinger Equation for a Hole Pair with the Same Flavor
Localized on a Rotating Skyrmion}

As we discussed before, a Skyrmion with the standard radial profile can only
localize $\alpha$-holes, while an anti-Skyrmion can localize $\beta$-holes.
Here we consider a general radial profile and study an $\alpha\alpha$ pair
localized on a rotating Skyrmion, or a $\beta\beta$ pair localized on a 
rotating anti-Skyrmion. It should be noted that the following symmetry analysis
even applies to an $\alpha\alpha$ pair localized on an anti-Skyrmion or
a $\beta\beta$ pair localized on a Skyrmion, but these configurations are
not energetically favorable.

If bound states of two holes of the same flavor $f$ localize on a rotating
Skyrmion, due to the Pauli principle, the holes cannot occupy the same state.
Let us therefore distinguish the holes by an unphysical label 1 or 2. If the
wave function is antisymmetric under the exchange of these labels, the Pauli
principle is satisfied. In this subsection we consider two holes of the same
flavor which is most relevant at low energies and relegate the localization of
two holes with different flavor to the Appendix.

The Hamiltonian describing two holes of the same flavor $f$ then reads
\begin{equation}
H = H^1 + H^2 + H^\gamma,
\end{equation}
where
\begin{eqnarray}
H^1&=&\left(\begin{array}{cccc}
H^f_{++} & 0 & H^f_{+-} & 0 \\ 0 & H^f_{++} & 0 & H^f_{+-} \\
H^f_{-+} & 0 & H^f_{--} & 0 \\ 0 & H^f_{-+} & 0 & H^f_{--} \end{array} \right),
\quad
H^2 = \left(\begin{array}{cccc}
H^f_{++} & H^f_{+-} & 0 & 0 \\ H^f_{-+} & H^f_{--} & 0 & 0 \\
0 & 0 & H^f_{++} & H^f_{+-} \\ 0 & 0 & H^f_{-+} & H^f_{--} \end{array} \right),
\nonumber \\
H^\gamma&=&\left(\begin{array}{cccc}
H^\gamma_{++++} & 0 & 0 & 0 \\
0 & H^\gamma_{+-+-} & 0 & 0 \\
0 & 0 & H^\gamma_{-+-+} & 0 \\
0 & 0 & 0 & H^\gamma_{----} \end{array} \right),
\end{eqnarray}
with
\begin{eqnarray}
\label{flavorHamiltonians}
&&H^\alpha_{++} = - \frac{1}{2 M'} (\p_i + i v^3_i(x))^2, \quad
H^\alpha_{+-} = \Lambda (iv_1^+(x) + v_2^+(x)), \nonumber \\
&&H^\alpha_{--} = - \frac{1}{2 M'} (\p_i - i v^3_i(x))^2, \quad
H^\alpha_{-+} = \Lambda (-iv_1^-(x) + v_2^-(x)), \nonumber \\
&&H^\beta_{++} = - \frac{1}{2 M'} (\p_i + i v^3_i(x))^2, \quad
H^\beta_{+-} = \Lambda (iv_1^+(x) - v_2^+(x)), \nonumber \\
&&H^\beta_{--} = - \frac{1}{2 M'} (\p_i - i v^3_i(x))^2, \quad
H^\beta_{-+} = \Lambda (-iv_1^-(x) - v_2^-(x)), \nonumber \\
&&H^\gamma_{++++} = - \frac{n^2}{2 {\cal D}(\rho) \rho^2}
\left(\p_\gamma + i n \frac{\Theta}{2 \pi} - i\frac{\sigma}{2}
(1-f(r_\alpha)) - i\frac{\sigma}{2}
(1-f(r_\beta))\right)^2, \nonumber \\
&&H^\gamma_{+-+-} = - \frac{n^2}{2 {\cal D}(\rho) \rho^2}
\left(\p_\gamma + i n \frac{\Theta}{2 \pi} - i\frac{\sigma}{2}
(1-f(r_\alpha)) + i\frac{\sigma}{2}
(1-f(r_\beta))\right)^2, \nonumber \\
&&H^\gamma_{-+-+} = - \frac{n^2}{2 {\cal D}(\rho) \rho^2}
\left(\p_\gamma + i n \frac{\Theta}{2 \pi} + i\frac{\sigma}{2}
(1-f(r_\alpha)) - i\frac{\sigma}{2}
(1-f(r_\beta))\right)^2, \nonumber \\
&&H^\gamma_{----} = - \frac{n^2}{2 {\cal D}(\rho) \rho^2}
\left(\p_\gamma + i n \frac{\Theta}{2 \pi} + i\frac{\sigma}{2}
(1-f(r_\alpha))+ i\frac{\sigma}{2} (1-f(r_\beta))\right)^2.
\end{eqnarray}

We first ignore the Pauli principle, and relegate the antisymmetrization of the
wave function in the labels 1 and 2 to the next section. This leads us to the
following ansatz for an energy eigenstate of two holes, which we have
distinguished by their labels,
\begin{eqnarray}
&&\hskip-1.5cm
\Psi^{ff}_{\sigma,m^1_+,m^1_-,m^2_+,m^2_-,m}(r_1,\chi_1,r_2,\chi_2,\gamma) =
\nonumber \\
&&\hskip-1.5cm\left(\begin{array}{c}
\psi^{ff}_{\sigma,m^1_+,m^1_-,m^2_+,m^2_-,m,++}(r_1,r_2)
\exp\left(i \sigma \left[m^1_+ \chi_1 + m^2_+ \chi_2
\right]\right) \exp(i \sigma (m - 1) \gamma) \\
\psi^{ff}_{\sigma,m^1_+,m^1_-,m^2_+,m^2_-,m,+-}(r_1,r_2)
\exp\left(i \sigma \left[m^1_+ \chi_1 + m^2_- \chi_2\right]\right)
\exp(i \sigma m \gamma) \\
\psi^{ff}_{\sigma,m^1_+,m^1_-,m^2_+,m^2_-,m,-+}(r_1,r_2)
\exp\left(i \sigma \left[m^1_- \chi_1 + m^2_+ \chi_2\right]\right)
\exp(i \sigma m \gamma) \\
\psi^{ff}_{\sigma,m^1_+,m^1_-,m^2_+,m^2_-,m,--}(r_1,r_2)
\exp\left(i \sigma \left[m^1_- \chi_1 + m^2_- \chi_2
\right]\right) \exp(i \sigma (m + 1) \gamma) \end{array}\right). \nonumber \\
\end{eqnarray}
This ansatz solves the Schr\"odinger equation, provided that
$m^{if}_- - m^{if}_+ = n + \sigma\sigma_f$, $i = 1, 2$. In this case, $m$ takes
integer values and the radial Schr\"odinger equation amounts to
\begin{equation}
\label{twoholeradialsame}
H_r \psi^{ff}_{\sigma,m^1_+,m^1_-,m^2_+,m^2_-,m}(r_1,r_2) =
E_{\sigma,m^1_+,m^1_-,m^2_+,m^2_-,m} \psi^{ff}_{\sigma,m^1_+,m^1_-,m^2_+,m^2_-,m}(r_1,r_2),
\end{equation}
with
\begin{equation}
\psi^{ff}_{\sigma,m^1_+,m^1_-,m^2_+,m^2_-,m}(r_1,r_2) =
\left(\begin{array}{c}
\psi^{ff}_{\sigma,m^1_+,m^1_-,m^2_+,m^2_-,m,++}(r_1,r_2) \\
\psi^{ff}_{\sigma,m^1_+,m^1_-,m^2_+,m^2_-,m,+-}(r_1,r_2) \\
\psi^{ff}_{\sigma,m^1_+,m^1_-,m^2_+,m^2_-,m,-+}(r_1,r_2) \\
\psi^{ff}_{\sigma,m^1_+,m^1_-,m^2_+,m^2_-,m,--}(r_1,r_2) \end{array}\right).
\end{equation}
The radial Hamiltonian is
\begin{equation}
H_r = H_r^1 + H_r^2 + H_r^\gamma,
\end{equation}
with
\begin{eqnarray}
H_r^1&=&\left(\begin{array}{cccc}
H^1_{r++} & 0 & H^1_{r+-} & 0 \\ 0 & H^1_{r++} & 0 & H^1_{r+-} \\
H^1_{r-+} & 0 & H^1_{r--} & 0 \\ 0 & H^1_{r-+} & 0 & H^1_{r--} \end{array} \right),
\nonumber \\
H_r^2&=&\left(\begin{array}{cccc}
H^2_{r++} & H^2_{r+-} & 0 & 0 \\ H^2_{r-+} & H^2_{r--} & 0 & 0 \\
0 & 0 & H^2_{r++} & H^2_{r+-} \\ 0 & 0 & H^2_{r-+} & H^2_{r--} \end{array} \right),
\nonumber \\
H_r^\gamma&=&\left(\begin{array}{cccc}
H^\gamma_{r++++} & 0 & 0 & 0 \\
0 & H^\gamma_{r+-+-} & 0 & 0 \\
0 & 0 & H^\gamma_{r-+-+} & 0 \\
0 & 0 & 0 & H^\gamma_{r----} \end{array} \right).
\end{eqnarray}
While the matrix elements related to the fermionic part of the radial
Hamiltonian are
\begin{eqnarray}
H^i_{r++}&=&- \frac{1}{2 M'}
\left[\p_{r_i}^2 + \frac{1}{r_i} \p_{r_i} - \frac{1}{r_i^2}
\left(m^{if}_+ + \frac{n \rho^{2n}}{r_i^{2n} + \rho^{2n}}
\right)^2\right],
\nonumber \\
H^i_{r+-}&=&H^i_{r-+} =
2 \Lambda \frac{n r_i^{n-1} \rho^{n}}{r_i^{2n} + \rho^{2n}}, \nonumber \\
H^i_{r--}&=&- \frac{1}{2 M'}
\left[\p_{r_i}^2 + \frac{1}{r_i} \p_{r_i} - \frac{1}{r_i^2}
\left(m^{if}_- - \frac{n \rho^{2n}}{r_i^{2n} + \rho^{2n}}\right)^2\right],
\end{eqnarray}
the contributions referring to the rotational Skyrmion read
\begin{eqnarray}
H^\gamma_{r++++}&=&\frac{n^2}{2 {\cal D}(\rho) \rho^2}
\left(m - 1 + \sigma n \frac{\Theta}{2 \pi} -
\frac{\rho^{2n}}{r_1^{2n} + \rho^{2n}} -
\frac{\rho^{2n}}{r_2^{2n} + \rho^{2n}}\right)^2, \nonumber \\
H^\gamma_{r+-+-}&=&\frac{n^2}{2 {\cal D}(\rho) \rho^2}
\left(m + \sigma n \frac{\Theta}{2 \pi} -
\frac{\rho^{2n}}{r_1^{2n} + \rho^{2n}} +
\frac{\rho^{2n}}{r_2^{2n} + \rho^{2n}}\right)^2, \nonumber \\
H^\gamma_{r-+-+}&=&\frac{n^2}{2 {\cal D}(\rho) \rho^2}
\left(m + \sigma n \frac{\Theta}{2 \pi} +
\frac{\rho^{2n}}{r_1^{2n} + \rho^{2n}} -
\frac{\rho^{2n}}{r_2^{2n} + \rho^{2n}}\right)^2, \nonumber \\
H^\gamma_{r----}&=&\frac{n^2}{2 {\cal D}(\rho) \rho^2}
\left(m + 1 + \sigma n \frac{\Theta}{2 \pi} +
\frac{\rho^{2n}}{r_1^{2n} + \rho^{2n}} +
\frac{\rho^{2n}}{r_2^{2n} + \rho^{2n}}\right)^2.
\end{eqnarray}

\subsection{Hole Pair of the Same Flavor Localized on a Skyrmion: Symmetry
Properties}

The spin operator $I$, which commutes with the two-hole Hamiltonian $H$,
amounts to
\begin{equation}
\label{spinoperator}
I = \left(\begin{array}{cccc}
- i \sigma \p_\gamma + \sigma n \frac{\Theta}{2 \pi} + 1 & 0 & 0 & 0 \\
0 & - i \sigma \p_\gamma + \sigma n \frac{\Theta}{2 \pi} & 0 & 0 \\
0 & 0 & - i \sigma \p_\gamma + \sigma n \frac{\Theta}{2 \pi} & 0 \\
0 & 0 & 0 & - i \sigma \p_\gamma + \sigma n \frac{\Theta}{2 \pi} - 1
\end{array} \right),
\end{equation}
such that
\begin{eqnarray}
&&I \Psi^{ff}_{\sigma,m^1_+,m^1_-,m^2_+,m^2_-,m}(r_1,\chi_1,r_2,\chi_2,\gamma) =
\nonumber \\
&&\left(m + \sigma n \frac{\Theta}{2 \pi}\right)
\Psi^{ff}_{\sigma,m^1_+,m^1_-,m^2_+,m^2_-,m}(r_1,\chi_1,r_2,\chi_2,\gamma).
\end{eqnarray}
Since $m$ takes integer values, at least for $\Theta = 0$, the state containing
two holes of the same flavor localized on a Skyrmion has integer spin as well.

Under the symmetries $D_i$, $O'$, and $R$ the two-hole wave function
\begin{equation}
\Psi^{ff}_{\sigma,n}(r_1,\chi_1,r_2,\chi_2,\gamma) =
\left(\begin{array}{c}
\Psi^{ff}_{\sigma,n,++}(r_1,\chi_1,r_2,\chi_2,\gamma) \\
\Psi^{ff}_{\sigma,n,+-}(r_1,\chi_1,r_2,\chi_2,\gamma) \\
\Psi^{ff}_{\sigma,n,-+}(r_1,\chi_1,r_2,\chi_2,\gamma) \\
\Psi^{ff}_{\sigma,n,--}(r_1,\chi_1,r_2,\chi_2,\gamma) \end{array}\right)
\end{equation}
transforms as
\begin{eqnarray}
&&^{D_i}\Psi^{ff}_{\sigma,n}(r_1,\chi_1,r_2,\chi_2,\gamma) =
\exp(2 i k^f a_i) \left(\begin{array}{c}
\Psi^{ff}_{\sigma,n,++}(r_1,\chi_1,r_2,\chi_2,\gamma) \\
\Psi^{ff}_{\sigma,n,+-}(r_1,\chi_1,r_2,\chi_2,\gamma) \\
\Psi^{ff}_{\sigma,n,-+}(r_1,\chi_1,r_2,\chi_2,\gamma) \\
\Psi^{ff}_{\sigma,n,--}(r_1,\chi_1,r_2,\chi_2,\gamma)
\end{array}\right), \nonumber \\
&&^{O'}\Psi^{ff}_{\sigma,n}(r_1,\chi_1,r_2,\chi_2,\gamma) =
\left(\begin{array}{c}
\exp(-\sigma_f \frac{4 \pi i}{3})\Psi^{ff}_{\sigma,n,--}(r_1,\chi_1 +
\frac{\pi}{3},
r_2,\chi_2 + \frac{\pi}{3},\gamma - n \frac{\pi}{3}) \\
- \Psi^{ff}_{\sigma,n,-+}(r_1,\chi_1 + \frac{\pi}{3},
r_2,\chi_2 + \frac{\pi}{3},\gamma - n \frac{\pi}{3}) \\
- \Psi^{ff}_{\sigma,n,+-}(r_1,\chi_1 + \frac{\pi}{3},
r_2,\chi_2 + \frac{\pi}{3},\gamma - n \frac{\pi}{3}) \\
\exp(\sigma_f \frac{4 \pi i}{3})\Psi^{ff}_{\sigma,n,++}(r_1,\chi_1 +
\frac{\pi}{3},
r_2,\chi_2 + \frac{\pi}{3},\gamma - n \frac{\pi}{3}) \end{array}\right),
\nonumber \\
&&^R\Psi^{ff}_{\sigma,n}(r_1,\chi_1,r_2,\chi_2,\gamma) =
\left(\begin{array}{c}
\Psi^{ff}_{\sigma,n,++}(r_1,-\chi_1,r_2,-\chi_2,-\gamma) \\
\Psi^{ff}_{\sigma,n,+-}(r_1,-\chi_1,r_2,-\chi_2,-\gamma) \\
\Psi^{ff}_{\sigma,n,-+}(r_1,-\chi_1,r_2,-\chi_2,-\gamma) \\
\Psi^{ff}_{\sigma,n,--}(r_1,-\chi_1,r_2,-\chi_2,-\gamma)
\end{array}\right).
\end{eqnarray}

Accordingly, the two-hole energy eigenstates transform as
\begin{eqnarray}
\label{symtwoholessame}
^{D_i}\Psi^{ff}_{\sigma,m^1_+,m^1_-,m^2_+,m^2_-,m}
(r_1,\chi_1,r_2,\chi_2,\gamma)\!\!\!\!&=&\!\!\!\!
\exp(2 i k^f a_i) \Psi^{ff}_{\sigma,m^1_+,m^1_-,m^2_+,m^2_-,m}
(r_1,\chi_1,r_2,\chi_2,\gamma), \nonumber \\
^{O'}\Psi^{\alpha\alpha}_{\sigma,m^1_+,m^1_-,m^2_+,m^2_-,m}
(r_1,\chi_1,r_2,\chi_2,\gamma)\!\!\!\!&=&\!\!\!\! - \exp\left(i
\sigma [m^1_+ + m^2_+ +n+\sigma - m n] \frac{\pi}{3}\right)
\nonumber \\
&\times&\!\!\!\!\Psi^{\beta\beta}_{-\sigma,-m^1_-,-m^1_+,-m^2_-,-m^2_+,-m}
(r_1,\chi_1,r_2,\chi_2,\gamma), \nonumber \\
^{O'}\Psi^{\beta\beta}_{\sigma,m^1_+,m^1_-,m^2_+,m^2_-,m}
(r_1,\chi_1,r_2,\chi_2,\gamma)\!\!\!\!&=&\!\!\!\! - \exp\left(i
\sigma [m^1_+ + m^2_+ +n-\sigma - m n] \frac{\pi}{3}\right)
\nonumber \\
&\times&\!\!\!\!\Psi^{\alpha\alpha}_{-\sigma,-m^1_-,-m^1_+,-m^2_-,-m^2_+,-m}
(r_1,\chi_1,r_2,\chi_2,\gamma), \nonumber \\
^R\Psi^{\alpha\alpha}_{\sigma,m^1_+,m^1_-,m^2_+,m^2_-,m}
(r_1,\chi_1,r_2,\chi_2,\gamma)\!\!\!\!&=&\!\!\!\!
\Psi^{\beta\beta}_{-\sigma,m^1_+,m^1_-,m^2_+,m^2_-,m}
(r_1,\chi_1,r_2,\chi_2,\gamma), \nonumber \\
^R\Psi^{\beta\beta}_{\sigma,m^1_+,m^1_-,m^2_+,m^2_-,m}
(r_1,\chi_1,r_2,\chi_2,\gamma)\!\!\!\!&=&\!\!\!\!
\Psi^{\alpha\alpha}_{-\sigma,m^1_+,m^1_-,m^2_+,m^2_-,m}
(r_1,\chi_1,r_2,\chi_2,\gamma).
\end{eqnarray}
Note that we have again used appropriate phase conventions for the radial
wave function $\psi_{\sigma,m^1_+,m^1_-,m^2_+,m^2_-,m}(r_1,r_2)$. Regarding the
rotation symmetries $O'$, we have used
\begin{eqnarray}
&&\psi^{ff}_{\sigma,m^1_+,m^1_-,m^2_+,m^2_-,m,--}(r_1,r_2) =
\psi^{f'f'}_{-\sigma,-m^1_-,-m^1_+,-m^2_-,-m^2_+,-m,++}(r_1,r_2), \nonumber \\
&&\psi^{ff}_{\sigma,m^1_+,m^1_-,m^2_+,m^2_-,m,-+}(r_1,r_2) =
\psi^{f'f'}_{-\sigma,-m^1_-,-m^1_+,-m^2_-,-m^2_+,-m,+-}(r_1,r_2), \nonumber \\
&&\psi^{ff}_{\sigma,m^1_+,m^1_-,m^2_+,m^2_-,m,+-}(r_1,r_2) =
\psi^{f'f'}_{-\sigma,-m^1_-,-m^1_+,-m^2_-,-m^2_+,-m,-+}(r_1,r_2), \nonumber \\
&&\psi^{ff}_{\sigma,m^1_+,m^1_-,m^2_+,m^2_-,m,++}(r_1,r_2) =
\psi^{f'f'}_{-\sigma,-m^1_-,-m^1_+,-m^2_-,-m^2_+,-m,--}(r_1,r_2).
\end{eqnarray}
These relations are a consequence of the symmetries of the radial Schr\"odinger
equation (\ref{twoholeradialsame}). With respect to the reflections $R$, we
have used
\begin{eqnarray}
\label{Rsymsame}
&&\psi^{ff}_{\sigma,m^1_+,m^1_-,m^2_+,m^2_-,m,++}(r_1,r_2) =
\psi^{f'f'}_{-\sigma,m^1_+,m^1_-,m^2_+,m^2_-,m,++}(r_1,r_2), \nonumber \\
&&\psi^{ff}_{\sigma,m^1_+,m^1_-,m^2_+,m^2_-,m,+-}(r_1,r_2) =
\psi^{f'f'}_{-\sigma,m^1_+,m^1_-,m^2_+,m^2_-,m,+-}(r_1,r_2), \nonumber \\
&&\psi^{ff}_{\sigma,m^1_+,m^1_-,m^2_+,m^2_-,m,-+}(r_1,r_2) =
\psi^{f'f'}_{-\sigma,m^1_+,m^1_-,m^2_+,m^2_-,m,-+}(r_1,r_2), \nonumber \\
&&\psi^{ff}_{\sigma,m^1_+,m^1_-,m^2_+,m^2_-,m,--}(r_1,r_2) =
\psi^{f'f'}_{-\sigma,m^1_+,m^1_-,m^2_+,m^2_-,m,--}(r_1,r_2).
\end{eqnarray}
The relations in Eq.~(\ref{Rsymsame}) are a consequence of the symmetries of
the radial Schr\"odinger equation (\ref{twoholeradialsame}) with $\Theta = 0$.
As before, in the case of $\Theta \neq 0$ or $\pi$, the reflection symmetry is
explicitly broken by the Hopf term.

We now incorporate the Pauli principle and explicitly antisymmetrize the wave
function in the artificial indices 1 and 2, by acting with the pair permutation
$P$,
\begin{equation}
^P\Psi^{ff}_{\sigma,n}(r_1,\chi_1,r_2,\chi_2,\gamma) =
\left(\begin{array}{c}
\Psi^{ff}_{\sigma,n,++}(r_2,\chi_2,r_1,\chi_1,\gamma) \\
\Psi^{ff}_{\sigma,n,-+}(r_2,\chi_2,r_1,\chi_1,\gamma) \\
\Psi^{ff}_{\sigma,n,+-}(r_2,\chi_2,r_1,\chi_1,\gamma) \\
\Psi^{ff}_{\sigma,n,--}(r_2,\chi_2,r_1,\chi_1,\gamma) \end{array}\right).
\end{equation}
For an energy eigenstate, we then have
\begin{equation}
^P\Psi^{ff}_{\sigma,m^1_+,m^1_-,m^2_+,m^2_-,m}(r_1,\chi_1,r_2,\chi_2,\gamma) =
\Psi^{ff}_{\sigma,m^2_+,m^2_-,m^1_+,m^1_-,m}(r_1,\chi_1,r_2,\chi_2,\gamma),
\end{equation}
assuming a symmetric radial wave function,
\begin{eqnarray}
&&\psi^{ff}_{\sigma,m^1_+,m^1_-,m^2_+,m^2_-,m,++}(r_2,r_1) =
\psi^{ff}_{\sigma,m^2_+,m^2_-,m^1_+,m^1_-,m,++}(r_1,r_2), \nonumber \\
&&\psi^{ff}_{\sigma,m^1_+,m^1_-,m^2_+,m^2_-,m,-+}(r_2,r_1) =
\psi^{ff}_{\sigma,m^2_+,m^2_-,m^1_+,m^1_-,m,+-}(r_1,r_2), \nonumber \\
&&\psi^{ff}_{\sigma,m^1_+,m^1_-,m^2_+,m^2_-,m,+-}(r_2,r_1) =
\psi^{ff}_{\sigma,m^2_+,m^2_-,m^1_+,m^1_-,m,-+}(r_1,r_2), \nonumber \\
&&\psi^{ff}_{\sigma,m^1_+,m^1_-,m^2_+,m^2_-,m,--}(r_2,r_1) =
\psi^{ff}_{\sigma,m^2_+,m^2_-,m^1_+,m^1_-,m,--}(r_1,r_2).
\end{eqnarray}
The properly antisymmetrized wave function then amounts to
\begin{equation}
\widetilde\Psi^{ff}_{\sigma,n}(r_1,\chi_1,r_2,\chi_2,\gamma) = \frac{1}{\sqrt{2}}
\left[\Psi^{ff}_{\sigma,n}(r_1,\chi_1,r_2,\chi_2,\gamma) -
^P\Psi^{ff}_{\sigma,n}(r_1,\chi_1,r_2,\chi_2,\gamma)\right],
\end{equation}
implying that for an energy eigenstate, we have
\begin{eqnarray}
&&\hspace{-2cm}
\widetilde\Psi^{ff}_{\sigma,m^1_+,m^1_-,m^2_+,m^2_-,m}(r_1,\chi_1,r_2,\chi_2,\gamma) =
\nonumber \\
&&\hspace{-2cm} \frac{1}{\sqrt{2}}
\left[\Psi^{ff}_{\sigma,m^1_+,m^1_-,m^2_+,m^2_-,m}(r_1,\chi_1,r_2,\chi_2,\gamma) -
\Psi^{ff}_{\sigma,m^2_+,m^2_-,m^1_+,m^1_-,m}(r_1,\chi_1,r_2,\chi_2,\gamma)\right].
\end{eqnarray}
Note that the two sets of quantum numbers $m^1_+, m^1_-$ and $m^2_+, m^2_-$ must
be different, in order to have a non-vanishing wave function. In the case of an
antisymmetric radial wave function, one could allow $m^1_+ = m^2_+$ and
$m^1_- = m^2_-$.

Using Eq.~(\ref{symtwoholessame}), the transformation properties of the
antisymmetrized two-hole energy eigenstates turn out to be
\begin{eqnarray}
\label{symtwoholessameanti}
^{D_i}\widetilde\Psi^{ff}_{\sigma,m^1_+,m^1_-,m^2_+,m^2_-,m}
(r_1,\chi_1,r_2,\chi_2,\gamma)\!\!\!\!&=&\!\!\!\!
\exp(2 i k^f a_i) \widetilde\Psi^{ff}_{\sigma,m^1_+,m^1_-,m^2_+,m^2_-,m}
(r_1,\chi_1,r_2,\chi_2,\gamma), \nonumber \\
^{O'}\widetilde\Psi^{\alpha\alpha}_{\sigma,m^1_+,m^1_-,m^2_+,m^2_-,m}
(r_1,\chi_1,r_2,\chi_2,\gamma)\!\!\!\!&=&\!\!\!\! - \exp\left(i
\sigma [m^1_+ + m^2_+ +n+\sigma - m n] \frac{\pi}{3}\right)
\nonumber \\
&\times&\!\!\!\!\widetilde\Psi^{\beta\beta}_{-\sigma,-m^1_-,-m^1_+,-m^2_-,-m^2_+,-m}
(r_1,\chi_1,r_2,\chi_2,\gamma), \nonumber \\
^{O'}\widetilde\Psi^{\beta\beta}_{\sigma,m^1_+,m^1_-,m^2_+,m^2_-,m}
(r_1,\chi_1,r_2,\chi_2,\gamma)\!\!\!\!&=&\!\!\!\! - \exp\left(i
\sigma [m^1_+ + m^2_+ +n-\sigma - m n] \frac{\pi}{3}\right)
\nonumber \\
&\times&\!\!\!\!\widetilde\Psi^{\alpha\alpha}_{-\sigma,-m^1_-,-m^1_+,-m^2_-,-m^2_+,-m}
(r_1,\chi_1,r_2,\chi_2,\gamma), \nonumber \\
^R\widetilde\Psi^{\alpha\alpha}_{\sigma,m^1_+,m^1_-,m^2_+,m^2_-,m}
(r_1,\chi_1,r_2,\chi_2,\gamma)\!\!\!\!&=&\!\!\!\!
\widetilde\Psi^{\beta\beta}_{-\sigma,m^1_+,m^1_-,m^2_+,m^2_-,m}
(r_1,\chi_1,r_2,\chi_2,\gamma), \nonumber \\
^R\widetilde\Psi^{\beta\beta}_{\sigma,m^1_+,m^1_-,m^2_+,m^2_-,m}
(r_1,\chi_1,r_2,\chi_2,\gamma)\!\!\!\!&=&\!\!\!\!
\widetilde\Psi^{\alpha\alpha}_{-\sigma,m^1_+,m^1_-,m^2_+,m^2_-,m}
(r_1,\chi_1,r_2,\chi_2,\gamma).
\end{eqnarray}

Finally, we take appropriate linear combinations of the states with flavors
$\alpha\alpha$ and $\beta\beta$ in order to obtain eigenstates of $O'$,
\begin{eqnarray}
& &
\widetilde\Psi^{\pm}_{\sigma,m^1_+,m^1_-,m^2_+,m^2_-,m}(r_1,\chi_1,r_2,\chi_2,\gamma)
= \frac{1}{\sqrt{2}}
\Big[\widetilde\Psi^{\alpha\alpha}_{\sigma,m^1_+,m^1_-,m^2_+,m^2_-,m}
(r_1,\chi_1,r_2,\chi_2,\gamma) \nonumber \\
& & \hspace{1.5cm} \pm
\widetilde\Psi^{\beta\beta}_{-\sigma,-m^1_-,-m^1_+,-m^2_-,-m^2_+,-m}
(r_1,\chi_1,r_2,\chi_2,\gamma) \Big], \nonumber \\
\end{eqnarray}
which transform as
\begin{eqnarray}
\label{symtwoholessamerot}
^{O'}\widetilde\Psi^{\pm}_{\sigma,m^1_+,m^1_-,m^2_+,m^2_-,m} (r_1,\chi_1,r_2,\chi_2
\gamma)\!\!\!\!&=&\!\!\!\! \mp\exp\left(i \sigma [m^1_+ + m^2_-  - m n]
\frac{\pi}{3}\right) \\
&\times&\!\!\!\!\widetilde\Psi^{\pm}_{\sigma,m^1_+,m^1_-,m^2_+,m^2_-,m} (r_1,\chi_1,
r_2,\chi_2,\gamma). \nonumber
\end{eqnarray}
We now systematically list all two-hole-Skyrmion wave functions ($n=1$, $m=0$)
that can be constructed with quantum numbers $m^i_{\pm}$ up to $\pm 2$. States
with $s$-wave symmetry are
\begin{eqnarray}
& & \widetilde\Psi^{+}_{-,2,2,1,1,0}(r_1,\chi_1,r_2,\chi_2,\gamma) , \nonumber \\
& & \widetilde\Psi^{+}_{-,1,1,2,2,0}(r_1,\chi_1,r_2,\chi_2,\gamma) , \nonumber \\
& & \widetilde\Psi^{+}_{-,-2,-2,-1,-1,0}(r_1,\chi_1,r_2,\chi_2,\gamma) , \nonumber
 \\
& & \widetilde\Psi^{+}_{-,-1,-1,-2,-2,0}(r_1,\chi_1,r_2,\chi_2,\gamma), \nonumber
\\
& & \widetilde\Psi^{-}_{+,-2,0,0,2,0}(r_1,\chi_1,r_2,\chi_2,\gamma) , \nonumber \\
& & \widetilde\Psi^{-}_{+,0,2,-2,0,0}(r_1,\chi_1,r_2,\chi_2,\gamma), \nonumber \\
& & \widetilde\Psi^{-}_{-,1,1,-1,-1,0}(r_1,\chi_1,r_2,\chi_2,\gamma) , \nonumber \\
& & \widetilde\Psi^{-}_{-,-1,-1,1,1,0}(r_1,\chi_1,r_2,\chi_2,\gamma) , \nonumber \\
& & \widetilde\Psi^{-}_{-,2,2,-2,-2,0}(r_1,\chi_1,r_2,\chi_2,\gamma) , \nonumber \\
& &\widetilde\Psi^{-}_{-,-2,-2,2,2,0}(r_1,\chi_1,r_2,\chi_2,\gamma) .
\end{eqnarray}
States with $p$-wave symmetry read
\begin{eqnarray}
& & \widetilde\Psi^{-}_{+,-1,1,-2,0,0}(r_1,\chi_1,r_2,\chi_2,\gamma) , \nonumber \\
& & \widetilde\Psi^{-}_{+,-2,0,-1,1,0}(r_1,\chi_1,r_2,\chi_2,\gamma) , \nonumber \\
& & \widetilde\Psi^{-}_{+,-1,1,0,2,0}(r_1,\chi_1,r_2,\chi_2,\gamma) , \nonumber \\
& & \widetilde\Psi^{-}_{+,0,2,-1,1,0}(r_1,\chi_1,r_2,\chi_2,\gamma) , \nonumber \\
& & \widetilde\Psi^{-}_{-,0,0,-1,-1,0}(r_1,\chi_1,r_2,\chi_2,\gamma) , \nonumber \\
& & \widetilde\Psi^{-}_{-,0,0,1,1,0}(r_1,\chi_1,r_2,\chi_2,\gamma) , \nonumber \\
& & \widetilde\Psi^{-}_{-,-1,-1,0,0,0}(r_1,\chi_1,r_2,\chi_2,\gamma) , \nonumber \\
& & \widetilde\Psi^{-}_{-,1,1,0,0,0}(r_1,\chi_1,r_2,\chi_2,\gamma) , \nonumber \\
& & \widetilde\Psi^{-}_{-,-1,-1,2,2,0}(r_1,\chi_1,r_2,\chi_2,\gamma) , \nonumber \\
& & \widetilde\Psi^{-}_{-,2,2,-1,-1,0}(r_1,\chi_1,r_2,\chi_2,\gamma) , \nonumber \\
& & \widetilde\Psi^{-}_{-,-2,-2,1,1,0}(r_1,\chi_1,r_2,\chi_2,\gamma) , \nonumber \\
& &
\widetilde\Psi^{-}_{-,1,1,-2,-2,0}(r_1,\chi_1,r_2,\chi_2,\gamma) .
\end{eqnarray}
States with $d$-wave symmetry correspond to
\begin{eqnarray}
& & \widetilde\Psi^{+}_{+,-1,1,-2,0,0}(r_1,\chi_1,r_2,\chi_2,\gamma) , \nonumber \\
& & \widetilde\Psi^{+}_{+,-2,0,-1,1,0}(r_1,\chi_1,r_2,\chi_2,\gamma) , \nonumber \\
& & \widetilde\Psi^{+}_{+,-1,1,0,2,0}(r_1,\chi_1,r_2,\chi_2,\gamma) , \nonumber \\
& & \widetilde\Psi^{+}_{+,0,2,-1,1,0}(r_1,\chi_1,r_2,\chi_2,\gamma) , \nonumber \\
& & \widetilde\Psi^{+}_{-,0,0,-1,-1,0}(r_1,\chi_1,r_2,\chi_2,\gamma) , \nonumber \\
& & \widetilde\Psi^{+}_{-,0,0,1,1,0}(r_1,\chi_1,r_2,\chi_2,\gamma) , \nonumber \\
& & \widetilde\Psi^{+}_{-,-1,-1,0,0,0}(r_1,\chi_1,r_2,\chi_2,\gamma) , \nonumber \\
& & \widetilde\Psi^{+}_{-,1,1,0,0,0}(r_1,\chi_1,r_2,\chi_2,\gamma) , \nonumber \\
& & \widetilde\Psi^{+}_{-,-1,-1,2,2,0}(r_1,\chi_1,r_2,\chi_2,\gamma) , \nonumber \\
& & \widetilde\Psi^{+}_{-,2,2,-1,-1,0}(r_1,\chi_1,r_2,\chi_2,\gamma) , \nonumber \\
& & \widetilde\Psi^{+}_{-,-2,-2,1,1,0}(r_1,\chi_1,r_2,\chi_2,\gamma) , \nonumber \\
& & \widetilde\Psi^{+}_{-,1,1,-2,-2,0}(r_1,\chi_1,r_2,\chi_2,\gamma) .
\end{eqnarray}
Finally, states with $f$-wave symmetry read
\begin{eqnarray}
& & \widetilde\Psi^{+}_{+,-2,0,0,2,0}(r_1,\chi_1,r_2,\chi_2,\gamma) , \nonumber \\
& & \widetilde\Psi^{+}_{+,0,2,-2,0,0}(r_1,\chi_1,r_2,\chi_2,\gamma), \nonumber \\
& & \widetilde\Psi^{+}_{-,1,1,-1,-1,0}(r_1,\chi_1,r_2,\chi_2,\gamma) , \nonumber \\
& & \widetilde\Psi^{+}_{-,-1,-1,1,1,0}(r_1,\chi_1,r_2,\chi_2,\gamma) , \nonumber \\
& & \widetilde\Psi^{+}_{-,2,2,-2,-2,0}(r_1,\chi_1,r_2,\chi_2,\gamma) , \nonumber \\
& & \widetilde\Psi^{+}_{-,-2,-2,2,2,0}(r_1,\chi_1,r_2,\chi_2,\gamma) , \nonumber \\
& & \widetilde\Psi^{-}_{-,2,2,1,1,0}(r_1,\chi_1,r_2,\chi_2,\gamma) , \nonumber \\
& & \widetilde\Psi^{-}_{-,1,1,2,2,0}(r_1,\chi_1,r_2,\chi_2,\gamma) , \nonumber \\
& & \widetilde\Psi^{-}_{-,-2,-2,-1,-1,0}(r_1,\chi_1,r_2,\chi_2,\gamma) , \nonumber
\\
& & \widetilde\Psi^{-}_{-,-1,-1,-2,-2,0}(r_1,\chi_1,r_2,\chi_2,\gamma) .
\end{eqnarray}
Interestingly, as we have shown in Ref.~\citep{Kae11}, if one restricts oneself
to the leading potentials that involve the $\Lambda$-term, one-magnon exchange
can only occur between two holes of different flavor. In the same flavor case,
this term merely leads to a contact interaction. Therefore there is nothing to
be compared with on the magnon-mediated bound-state side. It remains to be seen
whether magnon-mediated binding is indeed possible if one goes beyond the
leading $\Lambda$-term and how the symmetries of these bound states are related
to those corresponding to the localization of two holes on a Skyrmion. Note
that on the square lattice, the analogous $\Lambda$-term does lead to
one-magnon exchange between holes of different as well as of the same flavor.
Hence the formation of magnon-mediated two-hole bound states is possible in
either case, in contrast to the honeycomb lattice.

\section{Conclusions}

We have performed a careful symmetry analysis of the localization of doped holes
on a topological Skyrmion defect in the staggered magnetization order parameter
for an antiferromagnet on the honeycomb lattice. Our previous analysis for the
square lattice had shown that two holes residing in two different hole pockets
(and thus with different ``flavors'' $\alpha$ and $\beta$) naturally form a
bound pair, which may qualify as a preformed Cooper pair candidate for
high-temperature superconductivity. In the square lattice case, the most
attractive channel for hole pair formation turned out to have the same quantum
numbers as hole pairs bound by one-magnon exchange. In this work, we have seen
that on the honeycomb lattice the situation is qualitatively different than on
the square lattice. In particular, assuming the standard radial profile, a
Skyrmion (anti-Skyrmion) can only bind holes of flavor $\alpha$ ($\beta$).
Hence, unlike on the square lattice, the formation of $\alpha\beta$-pairs is
possible only for non-standard radial Skyrmion profiles. The question whether
such profiles are energetically favorable, requires detailed numerical
investigations of the dynamics. This goes beyond the symmetry analysis 
performed here, but provides an interesting topic for future studies. While
understanding the dynamical mechanism responsible for high-temperature
superconductivity remains extremely challenging, it seems promising to further
investigate hole pair localization on topological Skyrmion defects. Our symmetry
analysis, based on the systematic effective field theory for magnons and doped
holes, provides a solid theoretical basis for future investigations.

\section*{Acknowledgments}

C.\ P.\ H.\ and F.-J.\ J.\ thank the members of the Institute for Theoretical
Physics at Bern University for warm hospitality. The present project has
received funding from the Schweizerischer Na\-tio\-nal\-fonds as well as from
the European Research Council by means of the European Union's Seventh
Framework Programme (FP7/2007-2013)/ ERC grant agreement 339220.

\begin{appendix}

\section{Schr\"odinger Equation for Hole Pair with Different Flavor
Localized on a Rotating Skyr\-mi\-on}

In this appendix, we consider the analysis of bound states of two holes of
different flavor localized on the same (anti-)\-Skyr\-mi\-on. In contrast to two
holes of the same flavor, the localization of a hole of flavor $\alpha$ and a
second hole of flavor $\beta$ on the same (anti-)\-Skyr\-mi\-on is not excluded 
by the Pauli principle.

In the case of two holes of different flavor $\alpha$ and $\beta$, the
Hamiltonian takes the form
\begin{equation}
H = H^\alpha + H^\beta + H^\gamma.
\end{equation}
Here $H^\alpha$ and $H^\beta$ are the Hamiltonians for a hole of flavor $\alpha$
and $\beta$, respectively, given by
\begin{eqnarray}
H^\alpha&=&\left(\begin{array}{cccc}
H^\alpha_{++} & 0 & H^\alpha_{+-} & 0 \\
0 & H^\alpha_{++} & 0 & H^\alpha_{+-} \\
H^\alpha_{-+} & 0 & H^\alpha_{--} & 0 \\
0 & H^\alpha_{-+} & 0 & H^\alpha_{--} \end{array} \right), \quad
H^\beta = \left(\begin{array}{cccc}
H^\beta_{++} & H^\beta_{+-} & 0 & 0 \\
H^\beta_{-+} & H^\beta_{--} & 0 & 0 \\
0 & 0 & H^\beta_{++} & H^\beta_{+-} \\
0 & 0 & H^\beta_{-+} & H^\beta_{--} \end{array} \right), \nonumber \\
H^\gamma&=&\left(\begin{array}{cccc}
H^\gamma_{++++} & 0 & 0 & 0 \\
0 & H^\gamma_{+-+-} & 0 & 0 \\
0 & 0 & H^\gamma_{-+-+} & 0 \\
0 & 0 & 0 & H^\gamma_{----} \end{array} \right),
\end{eqnarray}
with $H^f_{\pm\pm}$ and $H^\gamma_{\pm\pm\pm\pm}$ given in
Eq.(\ref{flavorHamiltonians}).
For the two-hole energy eigenstate we make the ansatz
\begin{eqnarray}
&&\hskip-1.5cm
\Psi^{\alpha\beta}_{\sigma,m^\alpha_+,m^\alpha_-,m^\beta_+,m^\beta_-,m}
(r_\alpha,\chi_\alpha,r_\beta,\chi_\beta,\gamma) = \nonumber \\
&&\hskip-1.5cm\left(\begin{array}{c}
\psi^{\alpha \beta}_{\sigma,m^\alpha_+,m^\alpha_-,m^\beta_+,m^\beta_-,m,++}(r_\alpha,r_\beta)
\exp\left(i \sigma \left[m^\alpha_+ \chi_\alpha + m^\beta_+ \chi_\beta\right]
\right) \exp(i \sigma (m - 1) \gamma) \\
\psi^{\alpha \beta}_{\sigma,m^\alpha_+,m^\alpha_-,m^\beta_+,m^\beta_-,m,+-}(r_\alpha,r_\beta)
\exp\left(i \sigma \left[m^\alpha_+ \chi_\alpha + m^\beta_- \chi_\beta \right]
\right) \exp(i \sigma m \gamma) \\
\psi^{\alpha \beta}_{\sigma,m^\alpha_+,m^\alpha_-,m^\beta_+,m^\beta_-,m,-+}(r_\alpha,r_\beta)
\exp\left(i \sigma \left[m^\alpha_- \chi_\alpha + m^\beta_+ \chi_\beta \right]
\right) \exp(i \sigma m \gamma) \\
\psi^{\alpha \beta}_{\sigma,m^\alpha_+,m^\alpha_-,m^\beta_+,m^\beta_-,m,--}(r_\alpha,r_\beta)
\exp\left(i \sigma \left[m^\alpha_- \chi_\alpha + m^\beta_- \chi_\beta\right]
\right) \exp(i \sigma (m + 1) \gamma) \end{array}\right). \nonumber \\ \
\end{eqnarray}
Recall that the Schr\"odinger equation is only satisfied if
$m^f_- - m^f_+ = n + \sigma\sigma_f$. As for the Skyrmion with no holes
localized on it, here $m$ is again an integer. Accordingly, the radial
Schr\"odinger equation amounts to
\begin{equation}
\label{twoholeradial}
H_r \psi^{\alpha \beta}_{\sigma,m^\alpha_+,m^\alpha_-,m^\beta_+,m^\beta_-,m}
(r_\alpha,r_\beta) =
E_{\sigma,m^\alpha_+,m^\alpha_-,m^\beta_+,m^\beta_-,m}
\psi^{\alpha \beta}_{\sigma,m^\alpha_+,m^\alpha_-,m^\beta_+,m^\beta_-,m}(r_\alpha,r_\beta),
\end{equation}
with
\begin{equation}
\psi^{\alpha \beta}_{\sigma,m^\alpha_+,m^\alpha_-,m^\beta_+,m^\beta_-,m}(r_\alpha,r_\beta) =
\left(\begin{array}{c}
\psi^{\alpha \beta}_{\sigma,m^\alpha_+,m^\alpha_-,m^\beta_+,m^\beta_-,m,++}(r_\alpha,r_\beta)
\\
\psi^{\alpha \beta}_{\sigma,m^\alpha_+,m^\alpha_-,m^\beta_+,m^\beta_-,m,+-}(r_\alpha,r_\beta)
\\
\psi^{\alpha \beta}_{\sigma,m^\alpha_+,m^\alpha_-,m^\beta_+,m^\beta_-,m,-+}(r_\alpha,r_\beta)
\\
\psi^{\alpha \beta}_{\sigma,m^\alpha_+,m^\alpha_-,m^\beta_+,m^\beta_-,m,--}(r_\alpha,r_\beta)
\\
\end{array}\right).
\end{equation}
The various terms in the radial Hamiltonian
\begin{equation}
H_r = H_r^\alpha + H_r^\beta + H_r^\gamma,
\end{equation}
take the explicit form
\begin{eqnarray}
H_r^\alpha&=&\left(\begin{array}{cccc}
H^\alpha_{r++} & 0 & H^\alpha_{r+-} & 0 \\
0 & H^\alpha_{r++} & 0 & H^\alpha_{r+-} \\
H^\alpha_{r-+} & 0 & H^\alpha_{r--} & 0 \\
0 & H^\alpha_{r-+} & 0 & H^\alpha_{r--} \end{array} \right), \nonumber \\
H_r^\beta&=&\left(\begin{array}{cccc}
H^\beta_{r++} & H^\beta_{r+-} & 0 & 0 \\
H^\beta_{r-+} & H^\beta_{r--} & 0 & 0 \\
0 & 0 & H^\beta_{r++} & H^\beta_{r+-} \\
0 & 0 & H^\beta_{r-+} & H^\beta_{r--} \end{array} \right), \nonumber \\
H_r^\gamma&=&\left(\begin{array}{cccc}
H^\gamma_{r++++} & 0 & 0 & 0 \\
0 & H^\gamma_{r+-+-} & 0 & 0 \\
0 & 0 & H^\gamma_{r-+-+} & 0 \\
0 & 0 & 0 & H^\gamma_{r----} \end{array} \right).
\end{eqnarray}
The corresponding matrix elements of the fermionic part of the radial
Hamiltonian read
\begin{eqnarray}
H^f_{r++}&=&- \frac{1}{2 M'} \left[\p_{r_f}^2 + \frac{1}{r_f}
\p_{r_f} - \frac{1}{r_f^2} \left(m^f_+ + \frac{n }{2}(1-f(r_f))
\right)^2\right],
\nonumber \\
H^f_{r+-}&=&H^f_{r-+} = \frac{\Lambda}{2} \Big[
\frac{f'(r_f)}{\sqrt{1-f^2(r_f)}}
+ \frac{\sigma \sigma_f n}{r_f} \sqrt{1-f^2(r_f)} \Big] , \nonumber \\
H^f_{r--}&=&- \frac{1}{2 M'} \left[\p_{r_f}^2 + \frac{1}{r_f}
\p_{r_f} - \frac{1}{r_f^2} \left(m^f_- - \frac{n }{2}(1-f(r_f))
\right)^2\right],
\end{eqnarray}
while the contributions describing the rotating Skyrmion amount to
\begin{eqnarray}
H^\gamma_{r++++}&=&\frac{n^2}{2 {\cal D}(\rho) \rho^2} \left(m - 1 +
\sigma n \frac{\Theta}{2 \pi} -  \frac{1}{2}(1-f(r_\alpha)) -
 \frac{1}{2}(1-f(r_\beta))\right)^2, \nonumber \\
H^\gamma_{r+-+-}&=&\frac{n^2}{2 {\cal D}(\rho) \rho^2} \left(m +
\sigma n \frac{\Theta}{2 \pi} -  \frac{1}{2}(1-f(r_\alpha))
+ \frac{1}{2}(1-f(r_\beta))\right)^2, \nonumber \\
H^\gamma_{r-+-+}&=&\frac{n^2}{2 {\cal D}(\rho) \rho^2} \left(m +
\sigma n \frac{\Theta}{2 \pi}+  \frac{1}{2}(1-f(r_\alpha))
- \frac{1}{2}(1-f(r_\beta))\right)^2, \nonumber \\
H^\gamma_{r----}&=&\frac{n^2}{2 {\cal D}(\rho) \rho^2} \left(m + 1 +
\sigma n \frac{\Theta}{2 \pi} +  \frac{1}{2}(1-f(r_\alpha)) +
\frac{1}{2}(1-f(r_\beta))\right)^2.
\end{eqnarray}

\section{Hole Pair of Different Flavor Localized on a Skyrmion:
Symmetry Properties }

The spin operator $I$, Eq.~(\ref{spinoperator}), satisfies
\begin{eqnarray}
&&I \Psi^{\alpha\beta}_{\sigma,m^\alpha_+,m^\alpha_-,m^\beta_+,m^\beta_-,m}
(r_\alpha,\chi_\alpha,r_\beta,\chi_\beta,\gamma) = \nonumber \\
&&\left(m + \sigma n \frac{\Theta}{2 \pi}\right)
\Psi^{\alpha\beta}_{\sigma,m^\alpha_+,m^\alpha_-,m^\beta_+,m^\beta_-,m}
(r_\alpha,\chi_\alpha,r_\beta,\chi_\beta,\gamma).
\end{eqnarray}
Since $m$ takes integer values, at least for $\Theta = 0$, the state containing
two holes of different flavor localized on a Skyrmion has integer spin as well.

Under the symmetries $D_i$, $O'$, and $R$, the general two-hole wave function
\begin{equation}
\Psi^{\alpha \beta}_{\sigma,n}(r_\alpha,\chi_\alpha,r_\beta,\chi_\beta,\gamma) =
\left(\begin{array}{c}
\Psi^{\alpha \beta}_{\sigma,n,++}(r_\alpha,\chi_\alpha,r_\beta,\chi_\beta,\gamma) \\
\Psi^{\alpha \beta}_{\sigma,n,+-}(r_\alpha,\chi_\alpha,r_\beta,\chi_\beta,\gamma) \\
\Psi^{\alpha \beta}_{\sigma,n,-+}(r_\alpha,\chi_\alpha,r_\beta,\chi_\beta,\gamma) \\
\Psi^{\alpha \beta}_{\sigma,n,--}(r_\alpha,\chi_\alpha,r_\beta,\chi_\beta,\gamma)
\end{array}\right)
\end{equation}
transforms as
\begin{eqnarray}
&&^{D_i}\Psi^{\alpha \beta}_{\sigma,n}(r_\alpha,\chi_\alpha,r_\beta,\chi_\beta,\gamma)
= \exp(i(k^\alpha + k^\beta) a_i) \left(\begin{array}{c}
\Psi^{\alpha \beta}_{\sigma,n,++}(r_\alpha,\chi_\alpha,r_\beta,\chi_\beta,\gamma) \\
\Psi^{\alpha \beta}_{\sigma,n,+-}(r_\alpha,\chi_\alpha,r_\beta,\chi_\beta,\gamma) \\
\Psi^{\alpha \beta}_{\sigma,n,-+}(r_\alpha,\chi_\alpha,r_\beta,\chi_\beta,\gamma) \\
\Psi^{\alpha \beta}_{\sigma,n,--}(r_\alpha,\chi_\alpha,r_\beta,\chi_\beta,\gamma)
\end{array}\right), \nonumber \\
&&^{O'}\Psi^{\alpha \beta}_{\sigma,n}(r_\alpha,\chi_\alpha,r_\beta,\chi_\beta,\gamma) =
\left(\begin{array}{c}
\Psi^{\alpha \beta}_{\sigma,n,--}(r_\beta,\chi_\beta + \frac{\pi}{3},
r_\alpha,\chi_\alpha + \frac{\pi}{3},\gamma - n \frac{\pi}{3}) \\
- \exp \Big( \frac{4 \pi i}{3} \Big)
\Psi^{\alpha \beta}_{\sigma,n,+-}(r_\beta,\chi_\beta + \frac{\pi}{3},
r_\alpha,\chi_\alpha + \frac{\pi}{3},\gamma - n \frac{\pi}{3}) \\
- \exp \Big( -\frac{4 \pi i}{3} \Big) \Psi^{\alpha \beta}_{\sigma,n,-+}(r_\beta,
\chi_\beta
+ \frac{\pi}{3}, r_\alpha,\chi_\alpha + \frac{\pi}{3},\gamma - n \frac{\pi}{3})
\\
\Psi^{\alpha \beta}_{\sigma,n,++}(r_\beta,\chi_\beta + \frac{\pi}{3},
r_\alpha,\chi_\alpha + \frac{\pi}{3},\gamma - n \frac{\pi}{3})
\end{array}\right), \nonumber \\
&&^R\Psi^{\alpha \beta}_{\sigma,n}(r_\alpha,\chi_\alpha,r_\beta,\chi_\beta,\gamma) =
\left(\begin{array}{c}
\Psi^{\alpha \beta}_{\sigma,n,++}(r_\beta,-\chi_\beta,r_\alpha,-\chi_\alpha,-\gamma) \\
\Psi^{\alpha \beta}_{\sigma,n,-+}(r_\beta,-\chi_\beta,r_\alpha,-\chi_\alpha,-\gamma) \\
\Psi^{\alpha \beta}_{\sigma,n,+-}(r_\beta,-\chi_\beta,r_\alpha,-\chi_\alpha,-\gamma) \\
\Psi^{\alpha \beta}_{\sigma,n,--}(r_\beta,-\chi_\beta,r_\alpha,-\chi_\alpha,-\gamma)
\end{array}\right).
\end{eqnarray}

One readily derives the following transformation properties of the two-hole
energy eigenstates,
\begin{eqnarray}
\label{symtwoholes}
^{D_i}\Psi^{\alpha\beta}_{\sigma,m^\alpha_+,m^\alpha_-,m^\beta_+,m^\beta_-,m}
(r_\alpha,\chi_\alpha,r_\beta,\chi_\beta,\gamma)\!\!\!\!&=&\!\!\!\!
\Psi^{\alpha\beta}_{\sigma,m^\alpha_+,m^\alpha_-,m^\beta_+,m^\beta_-,m}
(r_\alpha,\chi_\alpha,r_\beta,\chi_\beta,\gamma), \nonumber \\
 \nonumber \\
^{O'}\Psi^{\alpha\beta}_{\sigma,m^\alpha_+,m^\alpha_-,m^\beta_+,m^\beta_-,m}
(r_\alpha,\chi_\alpha,r_\beta,\chi_\beta,\gamma)\!\!\!\!&=&\!\!\!\!
\exp\left(i \sigma [m^\alpha_+ + m^\beta_- +\sigma- m n]
\frac{\pi}{3}\right)
\nonumber \\
&\times&\!\!\!\!\Psi^{\alpha\beta}_{-\sigma,-m^\beta_-,-m^\beta_+,-m^\alpha_-,-m^\alpha_+,-m}
(r_\alpha,\chi_\alpha,r_\beta,\chi_\beta,\gamma), \nonumber \\
^R\Psi^{\alpha\beta}_{\sigma,m^\alpha_+,m^\alpha_-,m^\beta_+,m^\beta_-,m}
(r_\alpha,\chi_\alpha,r_\beta,\chi_\beta,\gamma)\!\!\!\!&=&\!\!\!\!
\Psi^{\alpha\beta}_{-\sigma,m^\beta_+,m^\beta_-,m^\alpha_+,m^\alpha_-,m}
(r_\alpha,\chi_\alpha,r_\beta,\chi_\beta,\gamma).
\end{eqnarray}
Note that we have assumed appropriate phase conventions for the radial wave
function $\psi_{\sigma,m^\alpha_+,m^\alpha_-,m^\beta_+,m^\beta_-,m}(r_\alpha,r_\beta)$,
which follow from the symmetries of the radial Schr\"odinger equation
(\ref{twoholeradial}). Explicitly, for the rotation symmetry $O'$ we have used
\begin{eqnarray}
&&\psi^{\alpha \beta}_{\sigma,m^\alpha_+,m^\alpha_-,m^\beta_+,m^\beta_-,m,--}
(r_\beta, r_\alpha) =
\psi^{\alpha \beta}_{-\sigma,-m^\beta_-,-m^\beta_+,-m^\alpha_-,-m^\alpha_+,-m,++}
(r_\alpha,r_\beta),
\nonumber \\
&&\psi^{\alpha \beta}_{\sigma,m^\alpha_+,m^\alpha_-,m^\beta_+,m^\beta_-,m,+-}
(r_\beta, r_\alpha) =
\psi^{\alpha \beta}_{-\sigma,-m^\beta_-,-m^\beta_+,-m^\alpha_-,-m^\alpha_+,-m,+-}
(r_\alpha,r_\beta),
\nonumber \\
&&\psi^{\alpha \beta}_{\sigma,m^\alpha_+,m^\alpha_-,m^\beta_+,m^\beta_-,m,-+}
(r_\beta, r_\alpha) =
\psi^{\alpha \beta}_{-\sigma,-m^\beta_-,-m^\beta_+,-m^\alpha_-,-m^\alpha_+,-m,-+}
(r_\alpha,r_\beta),
\nonumber \\
&&\psi^{\alpha \beta}_{\sigma,m^\alpha_+,m^\alpha_-,m^\beta_+,m^\beta_-,m,++}
(r_\beta, r_\alpha) =
\psi^{\alpha \beta}_{-\sigma,-m^\alpha_-,-m^\beta_+,-m^\alpha_-,-m^\alpha_+,-m,--}
(r_\alpha,r_\beta).
\end{eqnarray}
For the reflection symmetry $R$ we have used
\begin{eqnarray}
\label{Rsym}
&&\psi^{\alpha \beta}_{\sigma,m^\alpha_+,m^\alpha_-,m^\beta_+,m^\beta_-,m,++}
(r_\beta,r_\alpha) =
\psi^{\alpha \beta}_{-\sigma,m^\beta_+,m^\beta_-,m^\alpha_+,m^\alpha_-,m,++}
(r_\alpha,r_\beta), \nonumber \\
&&\psi^{\alpha \beta}_{\sigma,m^\alpha_+,m^\alpha_-,m^\beta_+,m^\beta_-,m,-+}
(r_\beta,r_\alpha) =
\psi^{\alpha \beta}_{-\sigma,m^\beta_+,m^\beta_-,m^\alpha_+,m^\alpha_-,m,+-}
(r_\alpha,r_\beta), \nonumber \\
&&\psi^{\alpha \beta}_{\sigma,m^\alpha_+,m^\alpha_-,m^\beta_+,m^\beta_-,m,+-}
(r_\beta,r_\alpha) =
\psi^{\alpha \beta}_{-\sigma,m^\beta_+,m^\beta_-,m^\alpha_+,m^\alpha_-,m,-+}
(r_\alpha,r_\beta), \nonumber \\
&&\psi^{\alpha \beta}_{\sigma,m^\alpha_+,m^\alpha_-,m^\beta_+,m^\beta_-,m,--}
(r_\beta,r_\alpha) =
\psi^{\alpha \beta}_{-\sigma,m^\beta_+,m^\beta_-,m^\alpha_+,m^\alpha_-,m,--}
(r_\alpha,r_\beta).
\end{eqnarray}
Again, the relations in Eq.(\ref{Rsym}) are a consequence of the symmetries of
the radial Schr\"odinger equation (\ref{twoholeradial}) for $\Theta = 0$. For
$\Theta \neq 0$ or $\pi$, the reflection symmetry is explicitly broken by the
Hopf term.

We now take appropriate linear combinations of Skyrmion and anti-Skyrmion
states which are eigenstates of the combined rotation $O'$,
\begin{eqnarray}
\label{LKdifferentFlavor}
& &
\Psi^{\pm}_{\sigma,m^{\alpha}_+,m^{\alpha}_-,m^{\beta}_+,m^{\beta}_-,m}(r_{\alpha},
\chi_{\alpha}, r_{\beta},\chi_{\beta},\gamma)
= \frac{1}{\sqrt{2}}
\Big[\Psi^{\alpha\beta}_{\sigma,m^{\alpha}_+,m^{\alpha}_-,m^{\beta}_+,m^{\beta}_-,m}
(r_{\alpha},\chi_{\alpha},r_{\beta},\chi_{\beta},\gamma) \nonumber \\
& & \hspace{1.5cm} \pm
\Psi^{\alpha\beta}_{-\sigma,-m^{\beta}_-,-m^{\beta}_+,-m^{\alpha}_-,-m^{\alpha}_+,-m}
(r_{\alpha},\chi_{\alpha},r_{\beta},\chi_{\beta},\gamma) \Big]. \nonumber \\
\end{eqnarray}
Under the various symmetries, these states transform as
\begin{eqnarray}
\label{TwoDiffHolesSymm}
^{D_i}\Psi^{\pm}_{\sigma,m^{\alpha}_+,m^{\alpha}_-,m^{\beta}_+,m^{\beta}_-,m}
(r_{\alpha},\chi_{\alpha},r_{\beta},\chi_{\beta},\gamma)\!\!\!\!&=&\!\!\!\!
\Psi^{\pm}_{\sigma,m^{\alpha}_+,m^{\alpha}_-,m^{\beta}_+,m^{\beta}_-,m}
(r_{\alpha},\chi_{\alpha},r_{\beta},\chi_{\beta},\gamma), \nonumber \\
^{O'}\Psi^{\pm}_{\sigma,m^{\alpha}_+,m^{\alpha}_-,m^{\beta}_+,m^{\beta}_-,m}
(r_{\alpha},\chi_{\alpha},r_{\beta},\chi_{\beta},\gamma)\!\!\!\!&=&\!\!\!\! \pm\exp
\left(i \sigma [m^{\alpha}_+ + m^{\beta}_- + \sigma - m n] \frac{\pi}{3}\right)
\nonumber \\
&\times&\!\!\!\!\Psi^{\pm}_{\sigma,m^{\alpha}_+,m^{\alpha}_-,m^{\beta}_+,
m^{\beta}_-,m}
(r_{\alpha},\chi_{\alpha},r_{\beta},\chi_{\beta},\gamma), \\
^R\Psi^{\pm}_{\sigma,m^{\alpha}_+,m^{\alpha}_-,m^{\beta}_+,m^{\beta}_-,m}
(r_{\alpha},\chi_{\alpha},r_{\beta},\chi_{\beta},\gamma)\!\!\!\!&=&\!\!\!\! \pm
\Psi^{\pm}_{\sigma,-m^{\alpha}_-,-m^{\alpha}_+,-m^{\beta}_-,-m^{\beta}_+,-m}
(r_{\alpha},\chi_{\alpha},r_{\beta},\chi_{\beta},\gamma). \nonumber
\end{eqnarray}

\section{Comparison with States of Two Holes Bound by One-Magnon Exchange}
\label{compar}

While we have considered the localization of two holes of different flavor on a 
Skyrmion in the previous section, it is instructive to compare this with the 
formation of two-hole bound states mediated by one-magnon exchange, discussed in
Ref.~\citep{Kae11}. In what follows, we first summarize these results.

In the rest frame, the Schr\"odinger equation describing the relative motion of
two holes with different flavors $\alpha$ and $\beta$, takes the form
\begin{equation}
\left(\begin{array}{cc} - \frac{1}{M'}\Delta & \gamma\frac{1}{\vec r^{\,2}}
\exp(-2i\varphi)
\\[0.2ex]
\gamma\frac{1}{\vec r^{\,2}}\exp(2i\varphi) &  - \frac{1}{M'} \Delta
\end{array} \right)
\left(\begin{array}{c} \Psi_1(\vec r \, ) \\
\Psi_2(\vec r \, ) \end{array}\right) = E
\left(\begin{array}{c} \Psi_1(\vec r \, ) \\
\Psi_2(\vec r \, ) \end{array}\right),
\end{equation}
where
\begin{equation}\label{gamma}
\gamma = \frac{\Lambda^2}{2\pi \rho_s}.
\end{equation}
The two probability amplitudes \mbox{$\Psi_1(\vec r \,)$} and
\mbox{$\Psi_2(\vec r \,)$} refer to the two spin-flavor combinations
\mbox{$\alpha_+\beta_-$} and \mbox{$\alpha_-\beta_+$}, respectively, with the
distance vector \mbox{$\vec r$} pointing from the \mbox{$\beta$} to the
\mbox{$\alpha$} hole. Note that the holes undergo a spin flip during the
one-magnon exchange process and that we are dealing with a two-component
Schr\"odinger equation describing the relative motion of the hole pair.

With the ansatz
\begin{equation}
\Psi_1(r,\varphi) = R_1(r) \exp (i m_1 \varphi), \qquad \Psi_2(r,\varphi) =
R_2(r) \exp (i m_2 \varphi),
\end{equation}
the radial and the angular part can be separated if the angular quantum numbers
obey the condition $m_2-m_1=2$. Using the parameter $\tilde m$,
\begin{equation}
m_1={\tilde m}-1, \qquad m_2={\tilde m}+1,
\end{equation}
the radial equations amount to
\begin{align}\label{radial_equations}
-\left(\frac{d^2}{d r^2}+\frac{1}{r}\frac{d}{d r} -\frac{1}{r^2}({\tilde m}-1)^2
\right)
R_1(r)+\gamma M'\frac{R_2(r)}{r^{2}} &= M' E R_1(r), \nonumber\\
-\left(\frac{d^2}{d r^2}+\frac{1}{r}\frac{d}{d r} -\frac{1}{r^2}({\tilde m}+1)^2
\right)
R_2(r)+\gamma M'\frac{R_1(r)}{r^{2}} &= M' E R_2(r).
\end{align}
In the particular case $\tilde m =0$, the set of equations reduces to
\begin{equation}
\label{radialequation}
\left[-\left(\frac{d^2}{d r^2}+\frac{1}{r}\frac{d}{d r}\right) +(1 -\gamma M')
\frac{1}{r^{2}}\right]R(r) = - M' |E| R(r),
\end{equation}
with $R(r)=R_1(r)-R_2(r)$, which is solved analytically by a modified Bessel
function
\begin{equation}
R(r) = A K_\nu \big( \sqrt{M' |E|} r \big), \qquad \nu = i \sqrt{\gamma M' -1}.
\end{equation}
We have modeled the short-range repulsion between the two holes with a
hard-core radius $r_0$ by requiring \mbox{$R(r_0)=0$} for \mbox{$r \leq r_0$}.
The potential in the radial equation (\ref{radialequation}) is negative and
thus attractive, provided that the low-energy constants satisfy the relation
\begin{equation}
1-\frac{M'\Lambda^2}{2\pi \rho_s} \leq 0.
\end{equation}
Accordingly, $\Lambda$ must be larger than the critical value
\begin{equation}
\Lambda_c = \sqrt{\frac{2\pi \rho_s}{M'}},
\end{equation}
in order for a magnon-mediated bound state to occur. The energy of the $n$-th
excited bound state amounts to
\begin{equation}
E_n \sim - \frac{1}{M' r_0^2} \exp\left(\frac{- 2 \pi n}{\sqrt{\gamma M' -1}}
\right).
\end{equation}

On the other hand, the angular part of the ground-state wave equation
($\tilde m = 0$) of two holes of flavors $\alpha$ and $\beta$ takes the simple
form
\begin{equation}
\Psi(r,\varphi) =
\left(\begin{array}{c} \Psi_1(\vec r \, ) \\
\Psi_2(\vec r \, ) \end{array}\right)
= R(r) \left(\begin{array}{c} \exp(-i\varphi \, ) \\
-\exp(i \varphi \, ) \end{array}\right) .
\end{equation}
While the ground state wave function is invariant under reflections $R$, shifts
$D_i$, and accidental continuous rotations $O(\gamma)$, under the combined
rotation $O'$, it picks up a sign,
\begin{equation}
^{O'}\Psi(r,\varphi) = - \Psi(r,\varphi).
\end{equation}
Although the probability distribution of the ground state seems to imply
$s$-wave symmetry (see Ref.~\citep{Kae11}, Fig.6), we are in fact dealing with
$f$-wave symmetry.

Let us now compare magnon-mediated hole binding with the localization of two
holes on a Skyrmion. In the case of a rotating Skyrmion with $n=1$ and $m=0$,
it is possible to construct a variety of two-hole-Skyrmion wave functions with
small quantum numbers. In the following we systematically enumerate $s$-,
$p$-, $d$- and $f$-wave states of the wave function, corresponding to quantum
numbers $m^{\alpha}_{\pm}, m^{\beta}_{\pm} $ up to $\pm 2$. We first consider the
linear combinations
\begin{equation}
\Psi^{+}_{\sigma,m^{\alpha}_+,m^{\alpha}_-,m^{\beta}_+,m^{\beta}_-,m}(r_{\alpha},
\chi_{\alpha}, r_{\beta},\chi_{\beta},\gamma)
\end{equation}
defined in Eq.~(\ref{LKdifferentFlavor}). States with $s$-wave symmetry are
\begin{eqnarray}
& & \Psi^{+}_{+,-1,1,0,0,0}(r_\alpha,\chi_\alpha,r_\beta,\chi_\beta,
\gamma) , \nonumber \\
& & \Psi^{+}_{+,0,2,-1,-1,0}(r_\alpha,\chi_\alpha,r_\beta,\chi_\beta,
\gamma), \nonumber \\
& & \Psi^{+}_{+,-2,0,1,1,0}(r_\alpha,\chi_\alpha,r_\beta,\chi_\beta,
\gamma) , \nonumber \\
& & \Psi^{+}_{-,0,0,-1,1,0}(r_\alpha,\chi_\alpha,r_\beta,\chi_\beta,
\gamma) , \nonumber \\
& & \Psi^{+}_{-,1,1,-2,0,0}(r_\alpha,\chi_\alpha,r_\beta,\chi_\beta,
\gamma) , \nonumber \\
& & \Psi^{+}_{-,-1,-1,0,2,0}(r_\alpha,\chi_\alpha,r_\beta,\chi_\beta,
\gamma) .
\end{eqnarray}
States with $p$-wave symmetry read
\begin{eqnarray}
& & \Psi^{+}_{+,-1,1,1,1,0}(r_\alpha,\chi_\alpha,r_\beta,\chi_\beta,
\gamma) ,
\nonumber \\
& & \Psi^{+}_{+,-1,1,-1,-1,0}(r_\alpha,\chi_\alpha,r_\beta,\chi_\beta
\gamma) ,
\nonumber \\
& & \Psi^{+}_{+,0,2,0,0,0}(r_\alpha,\chi_\alpha,r_\beta,\chi_\beta,
\gamma) ,
\nonumber \\
& & \Psi^{+}_{+,-2,0,0,0,0}(r_\alpha,\chi_\alpha,r_\beta,\chi_\beta,
\gamma) ,
\nonumber \\
& & \Psi^{+}_{+,0,2,-2,-2,0}(r_\alpha,\chi_\alpha,r_\beta,\chi_\beta,
\gamma)
,
\nonumber \\
& &\Psi^{+}_{+,-2,0,2,2,0}(r_\alpha,\chi_\alpha,r_\beta,\chi_\beta,
\gamma) ,
\nonumber \\
& &\Psi^{+}_{-,0,0,-2,0,0}(r_\alpha,\chi_\alpha,r_\beta,\chi_\beta,
\gamma) ,
\nonumber \\
& & \Psi^{+}_{-,0,0,0,2,0}(r_\alpha,\chi_\alpha,r_\beta,\chi_\beta,
\gamma) ,
\nonumber \\
& & \Psi^{+}_{-,1,1,-1,1,0}(r_\alpha,\chi_\alpha,r_\beta,\chi_\beta,
\gamma) ,
\nonumber \\
& & \Psi^{+}_{-,-1,-1,-1,1,0}(r_\alpha,\chi_\alpha,r_\beta,\chi_\beta,
\gamma) ,
\nonumber \\
& & \Psi^{+}_{-,2,2,-2,0,0}(r_\alpha,\chi_\alpha,r_\beta,\chi_\beta,
\gamma) ,
\nonumber \\
& & \Psi^{+}_{-,-2,-2,0,2,0}(r_\alpha,\chi_\alpha,r_\beta,\chi_\beta,
\gamma)
 .
\end{eqnarray}
States with $d$-wave symmetry correspond to
\begin{eqnarray}
& & \Psi^{+}_{+,0,2,1,1,0}(r_\alpha,\chi_\alpha,r_\beta,\chi_\beta,
\gamma) ,
\nonumber \\
& & \Psi^{+}_{+,-2,0,-1,-1,0}(r_\alpha,\chi_\alpha,r_\beta,\chi_\beta,
\gamma) ,
\nonumber \\
& & \Psi^{+}_{+,-1,1,2,2,0}(r_\alpha,\chi_\alpha,r_\beta,\chi_\beta,
\gamma) ,
\nonumber \\
& & \Psi^{+}_{+,-1,1,-2,-2,0}(r_\alpha,\chi_\alpha,r_\beta,\chi_\beta,
\gamma) ,
\nonumber \\
& & \Psi^{+}_{-,1,1,0,2,0}(r_\alpha,\chi_\alpha,r_\beta,\chi_\beta,
\gamma) ,
\nonumber \\
& & \Psi^{+}_{-,-1,-1,-2,0,0}(r_\alpha,\chi_\alpha,r_\beta,\chi_\beta,
\gamma) ,
\nonumber \\
& & \Psi^{+}_{-,2,2,-1,1,0}(r_\alpha,\chi_\alpha,r_\beta,\chi_\beta,
\gamma) ,
\nonumber \\
& & \Psi^{+}_{-,-2,-2,-1,1,0}(r_\alpha,\chi_\alpha,r_\beta,\chi_\beta,
\gamma) .
\end{eqnarray}
Finally, states with $f$-wave symmetry read
\begin{eqnarray}
& & \Psi^{+}_{+,0,2,2,2,0}(r_\alpha,\chi_\alpha,r_\beta,\chi_\beta,
\gamma) ,
\nonumber \\
& & \Psi^{+}_{+,-2,0,-2,-2,0}(r_\alpha,\chi_\alpha,r_\beta,\chi_\beta,
\gamma) ,
\nonumber \\
& & \Psi^{+}_{-,2,2,0,2,0}(r_\alpha,\chi_\alpha,r_\beta,\chi_\beta,
\gamma) ,
\nonumber \\
& & \Psi^{+}_{-,-2,-2,-2,0,0}(r_\alpha,\chi_\alpha,r_\beta,\chi_\beta,
\gamma) .
\end{eqnarray}
As far as the other linear combinations
\begin{equation}
\Psi^{-}_{\sigma,m^{\alpha}_+,m^{\alpha}_-,m^{\beta}_+,m^{\beta}_-,m}(r_{\alpha},
\chi_{\alpha}, r_{\beta},\chi_{\beta},\gamma)
\end{equation}
are concerned, due to the extra minus sign under the transformation $O'$ in
Eq.~(\ref{TwoDiffHolesSymm}), the above classification is still valid, provided
one makes the replacements $s \Longleftrightarrow f$ and
$p \Longleftrightarrow d$.

Note that the $f$-wave states differ in their transformation properties under
$R$ with respect to the magnon-mediated two-hole ground states
$\Psi(r,\varphi)$ which are invariant under $R$. Accordingly, in contrast to
the square lattice case, there is no one-to-one correspondence between these
magnon-mediated two-hole bound states and the $f$-wave states formed by two
holes localized on a Skyrmion. Still, one of the candidate two-hole-Skyrmion
ground states, and one of the candidate two-hole-anti-Skyrmion ground states
(i.e., those with smallest quantum numbers) correspond to the wave functions
$\Psi^{-}_{+,-1,1,0,0,0}(r_\alpha,\chi_\alpha,r_\beta,\chi_\beta,\gamma)$
and $\Psi^{-}_{-,0,0,-1,1,0}(r_\alpha,\chi_\alpha,r_\beta,\chi_\beta,
\gamma)$, which both have $f$-wave symmetry. It should be pointed out again
that the pairing symmetry in the dehydrated version of 
Na$_2$CoO$_2 \times y$H$_2$O --- an experimental realization of a hole-doped 
honeycomb lattice antiferromagnet --- indeed appears to be $f$-wave 
\citep{MJ05}.

\end{appendix}


\begin{thebibliography}{10}

\bibitem{Bru06}
C.\ Br\"ugger, F.\ K\"ampfer, M.\ Moser, M.\ Pepe, and U.-J.\ Wiese,
Phys.\ Rev.\ B \textbf{74}, 224432 (2006).

\bibitem{Kae11}
F.\ K\"ampfer, B.\ Bessire, M.\ Wirz, C.\ P.\ Hofmann, F.-J.\ Jiang, and
U.-J.\ Wiese, Phys.\ Rev.\ B \textbf{85}, 075123 (2012).

\bibitem{Gas88}
J.\ Gasser, M.\ E.\ Sainio, and A.\ Svarc, Nucl.\ Phys.\ B \textbf{307}, 779
(1988).

\bibitem{Jen91}
E.\ Jenkins and A.\ Manohar, Phys.\ Lett.\ B \textbf{255}, 558 (1991).

\bibitem{Ber92}
V.\ Bernard, N.\ Kaiser, J.\ Kambor, and U.-G.\ Meissner, Nucl.\ Phys.\
B \textbf{388}, 315 (1992).

\bibitem{Bec99}
T.\ Becher and H.\ Leutwyler, Eur.\ Phys.\ J.\ C \textbf{9}, 643 (1999).

\bibitem{Kae05}
F.\ K\"ampfer, M.\ Moser, and U.-J.\ Wiese, Nucl.\ Phys.\ B \textbf{729}, 317
(2005).

\bibitem{Bru05}
C.\ Br\"ugger, F.\ K\"ampfer, M.\ Pepe, and U.-J.\ Wiese, Eur.\ Phys.\ J.\
B \textbf{53}, 433 (2006).

\bibitem{Bru07}
C.\ Br\"ugger, C.\ P.\ Hofmann, F.\ K\"ampfer, M.\ Pepe, and U.-J.\ Wiese,
Phys.\ Rev.\ B \textbf{75}, 014421 (2007).

\bibitem{Bru07a}
C.\ Br\"ugger, C.\ P.\ Hofmann, F.\ K\"ampfer, M.\ Moser, M.\ Pepe, and
U.-J.\ Wiese, Phys.\ Rev.\ B \textbf{75}, 214405 (2007).

\bibitem{Jia09}
F.-J.\ Jiang, F.\ K\"ampfer, C.\ P.\ Hofmann, and U.-J.\ Wiese,
Eur.\ Phys.\ J.\ B \textbf{69}, 473 (2009).

\bibitem{VHJW12}
N.\ D.\ Vlasii, C.\ P.\ Hofmann, F.-J.\ Jiang, and U.-J.\ Wiese,
Phys.\ Rev.\ B \textbf{86}, 155113 (2012).

\bibitem{Hal88}
F.\ D.\ M.\ Haldane, Phys.\ Rev.\ Lett.\ \textbf{61}, 1029 (1988).

\bibitem{Rea89}
N.\ Read and S.\ Sachdev, Phys.\ Rev.\ Lett.\ \textbf{62}, 1694 (1989);
Nucl.\ Phys.\ B \textbf{316}, 609 (1989).

\bibitem{Shr90}
B.\ I.\ Shraiman and E.\ D.\ Siggia, Phys.\ Rev.\ B \textbf{42}, 2485 (1990).

\bibitem{Goo91}
R.\ J.\ Gooding, Phys.\ Rev.\ Lett.\ \textbf{66}, 2266 (1991).

\bibitem{Goo93}
R.\ J.\ Gooding and A.\ Mailhot, Phys.\ Rev.\ B \textbf{48}, 6132 (1993).

\bibitem{Haa96}
S.\ Haas, F.-C.\ Zhang, F.\ Mila, and T.\ M.\ Rice,
Phys.\ Rev.\ Lett.\ \textbf{77}, 3021 (1996).

\bibitem{Mar01}
E.\ C.\ Marino and M.\ B.\ Silva Neto, Phys.\ Rev.\ B \textbf{64}, 092511
(2001).

\bibitem{Mot03}
O.\ I.\ Motrunich and A.\ Vishwanath, Phys.\ Rev.\ B \textbf{70}, 075104 (2004).

\bibitem{Sen04}
T.\ Sentil, A.\ Vishwanath, L.\ Balents, S.\ Sachdev, and M.\ P.\ A.\ Fisher,
Science \textbf{303}, 1490 (2004); Phys.\ Rev.\ B \textbf{70}, 144407 (2004).

\bibitem{Bae04}
O.\ B\"ar, M.\ Imboden, and U.-J.\ Wiese, Nucl.\ Phys.\ B \textbf{686}, 347
(2004).

\bibitem{Wie05}
U.-J.\ Wiese, Nucl.\ Phys.\ Proc.\ Suppl.\ \textbf{141}, 143 (2005).

\bibitem{Mor05}
T.\ Morinari, Phys.\ Rev.\ B \textbf{72}, 104502 (2005).

\bibitem{Fu10}
L.\ Fu, S.\ Sachdev, and C.\ Xu, Phys.\ Rev.\ B \textbf{83}, 165123 (2011).

\bibitem{Rai11}
I.\ Raicevic, D.\ Popovic, C.\ Panagopoulos, L.\ Benfatto, M.\ B.\ Silva Neto,
E.\ S.\ Choi, and T.\ Sasagawa, Phys.\ Rev.\ Lett.\ \textbf{106}, 227206 (2011).

\bibitem{Bas11}
G.\ Baskaran, arXiv:1108.3562.

\bibitem{Ver91}
J.\ A.\ Verg\'es, E.\ Louis, P.\ S.\ Lohmdahl, F.\ Guinea, and A.\ R.\ Bishop,
Phys.\ Rev.\ B \textbf{43}, 6099 (1991).

\bibitem{Sei98}
G.\ Seibold, Phys.\ Rev.\ B \textbf{58}, 15520 (1998).

\bibitem{Ber99}
M.\ Berciu and S.\ John, Phys.\ Rev.\ B \textbf{59}, 15143 (1999).

\bibitem{Tim00}
C.\ Timm and K.\ H.\ Bennemann, Phys.\ Rev.\ Lett.\ \textbf{84}, 4994 (2000).

\bibitem{MJ05}
I.\ I.\ Mazin and M.\ D.\ Johannes, Nat.\ Phys.\ \textbf{1}, 91 (2005).

\bibitem{Hon08}
C.\ Honercamp, Phys.\ Rev.\ Lett.\ \textbf{100}, 146404 (2008).

\bibitem{Lee10}
W.-C.\ Lee, C.\ Wu, and S.\ Das Sarma, Phys.\ Rev.\ A \textbf{82}, 053611 
(2010).

\bibitem{Gan13}
R.\ Ganesh, S.\ Nishimoto, and J.\ van den Brink, Phys.\ Rev.\ B \textbf{87},
054413 (2013).

\bibitem{Wen88}
X.\ G.\ Wen and A.\ Zee, Phys.\ Rev.\ Lett.\ \textbf{61}, 1025 (1988).

\bibitem{Dom88}
T.\ Dombre and N.\ Read, Phys.\ Rev.\ B \textbf{38,} 7181 (1988).

\bibitem{Fra88}
E.\ Fradkin and M.\ Stone, Phys.\ Rev.\ B \textbf{38}, 7215 (1988).

\bibitem{Jia08}
F.-J.\ Jiang, F.\ K\"ampfer, M.\ Nyfeler, and U.-J.\ Wiese, Phys.\ Rev.\ B
\textbf{78}, 214406 (2008). 

\end{thebibliography}
\end{document}